\documentclass[twocolumn]{aastex631}
\shorttitle{On the dwarf irregular NGC~6822}
\shortauthors{Tantalo et al.}

\graphicspath{{./}{figures/}}

\begin{document}

\title{On the dwarf irregular galaxy NGC~6822. I. Young, intermediate and old stellar populations.}%\footnote{Released on ...}}

\correspondingauthor{Maria Tantalo}
\email{maria.tantalo@students.uniroma2.eu, maria.tantalo@inaf.it}

\author[0000-0002-6829-6704]{Maria Tantalo}
\affiliation{Dipartimento di Fisica, Università di Roma Tor Vergata, via della Ricerca Scientifica 1, 00133 Roma, Italy}
\affiliation{INAF - Osservatorio Astronomico di Roma, via Frascati 33, 00078 Monte Porzio Catone, Italy}

\author{Massimo Dall'Ora}
\affiliation{INAF - Osservatorio Astronomico di Capodimonte, Salita Moiariello 16, 80131 Napoli, Italy}

\author{Giuseppe Bono}
\affiliation{Dipartimento di Fisica, Università di Roma Tor Vergata, via della Ricerca Scientifica 1, 00133 Roma, Italy}
\affiliation{INAF - Osservatorio Astronomico di Roma, via Frascati 33, 00078 Monte Porzio Catone, Italy}

\author{Peter B. Stetson}
\affiliation{Herzberg Astronomy and Astrophysics, National Research Council, 5071 West Saanich Road, Victoria, British Columbia V9E 2E7, Canada}

\author{Michele Fabrizio}
\affiliation{INAF - Osservatorio Astronomico di Roma, via Frascati 33, 00078 Monte Porzio Catone, Italy}
\affiliation{Space Science Data Center, via del Politecnico snc, 00133 Roma, Italy}

\author{Ivan Ferraro}
\affiliation{INAF - Osservatorio Astronomico di Roma, via Frascati 33, 00078 Monte Porzio Catone, Italy}

\author{Mario Nonino}
\affiliation{INAF – Osservatorio Astronomico di Trieste, Via G. B. Tiepolo 11, 34143 Trieste, Italy}

\author{Vittorio F. Braga}
\affiliation{INAF - Osservatorio Astronomico di Roma, via Frascati 33, 00078 Monte Porzio Catone, Italy}
\affiliation{Space Science Data Center, via del Politecnico snc, 00133 Roma, Italy}
\affiliation{Instituto de Astrofísica de Canarias, Calle Via Lactea s/n, E-38205 La Laguna, Tenerife, Spain}

\author{Ronaldo da Silva}
\affiliation{INAF - Osservatorio Astronomico di Roma, via Frascati 33, 00078 Monte Porzio Catone, Italy}
\affiliation{Space Science Data Center, via del Politecnico snc, 00133 Roma, Italy}

\author{Giuliana Fiorentino}
\affiliation{INAF - Osservatorio Astronomico di Roma, via Frascati 33, 00078 Monte Porzio Catone, Italy}

\author{Giacinto Iannicola}
\affiliation{INAF - Osservatorio Astronomico di Roma, via Frascati 33, 00078 Monte Porzio Catone, Italy}

\author{Massimo Marengo}
\affiliation{Department of Physics and Astronomy, Iowa State University, Ames, IA 50011, USA}

\author{Matteo Monelli}
\affiliation{Instituto de Astrofísica de Canarias, Calle Via Lactea s/n, E-38205 La Laguna, Tenerife, Spain}
\affiliation{Departamento de Astrofísica, Universidad de La Laguna (ULL), E-38200, La Laguna, Tenerife, Spain}

\author{Joseph P. Mullen}
\affiliation{Department of Physics and Astronomy, Iowa State University, Ames, IA 50011, USA}

\author{Adriano Pietrinferni}
\affiliation{INAF - Osservatorio Astronomico d’Abruzzo, via M. Maggini, s/n, I-64100, Teramo, Italy}

\author{Maurizio Salaris}
\affiliation{Astrophysics Research Institute, Liverpool John Moores University, 146 Brownlow Hill, Liverpool L3 5RF, United Kingdom}
\affiliation{INAF - Osservatorio Astronomico d’Abruzzo, via M. Maggini, s/n, I-64100, Teramo, Italy}

\begin{abstract}

We present accurate and deep multi-band ($g,r,i$) photometry of the
Local Group dwarf irregular galaxy NGC~6822. The images were collected with
wide field cameras at 2m/4m- (INT,CTIO,CFHT) and 8m-class telescopes (SUBARU)
covering a 2~square degrees FoV across the center of the galaxy. We performed
PSF photometry of $\approx$7,000~CCD images and the final catalog includes 
more than 1~million objects. We developed a new approach to identify
candidate field and galaxy stars, and performed a new estimate of the galaxy
center by using old stellar tracers finding that it differs by 1.15 (RA) and
1.53 (DEC) arcmin from previous estimates.
We also found that young (Main Sequence, Red Supergiants), intermediate (Red Clump,
Asymptotic Giant Branch [AGB]) and old (Red Giant Branch [RGB]) stars display different
radial distributions. Old stellar population is spherically distributed and
extends to radial distances larger than previously estimated ($\sim$1~degree).
The young population shows a well defined bar and a disk-like distribution, as
suggested by radio measurements, that is off-center compared with old population.
We discuss pros and cons of the different diagnostics adopted to identify 
AGB stars and develop new ones based on optical-NIR-MIR color-color diagrams (CCDs)
to characterize Oxygen and Carbon (C) rich stars.
We found a mean population ratio between Carbon and M-type (C/M) stars of
0.67$\pm$0.08 (optical/NIR/MIR) and we used the observed C/M ratio
with empirical C/M-metallicity relations to estimate a mean iron abundance of 
[Fe/H]$\sim$-1.25 ($\sigma$=0.04 dex) that agrees quite well with literature estimates.

\end{abstract}


\defcitealias{Kacharov12}{K12}
\defcitealias{Letarte02}{L02}
\defcitealias{Sibbons12}{S12}
\defcitealias{Sibbons15}{S15}

%%%%%%%%%%%%%%%%%%%%%%%%%%%%%%%%%%%%%%%%%%%%%%%%%%%%%%%%%%%%%%%%%%%%%%%%%%%%%%%%%%%%%
\section{Introduction} \label{sec:intro}
%%%%%%%%%%%%%%%%%%%%%%%%%%%%%%%%%%%%%%%%%%%%%%%%%%%%%%%%%%%%%%%%%%%%%%%%%%%%%%%%%%%%%

Stellar systems hosting stellar populations ranging from very young (a few Myr) 
main sequence stars to old (t$>$10 Gyr) low-mass stars are rare in the Local Group (LG). 
The most prominent systems, but the three large LG galaxies (Milky Way [MW], M31, M33), 
are the Magellanic Clouds (MCs). However, we still lack firm theoretical and 
empirical constraints of their mutual dynamical interaction and of their interaction 
with the MW. This means that the star formation episodes and their chemical enrichment 
histories might have been either affected or driven by the environment. 
The three next stellar systems sharing similar metallicity distributions
and hosting Wolf Rayet stars are the three LG dwarf irregulars IC~10, IC~1613 and
NGC~6822 \citep{Neugent19}. These three systems are considered
unique analogs of star forming galaxies observed at high redshift and
populating the universe at the peak of cosmic star formation history \citep{Stott13, Du20}.
       
NGC~6822 brings forward several key properties that make this stellar system particularly 
interesting (key positional, structural and photometric properties are listed in Table~\ref{tab:properties}). 
a) The position on the sky \citep[$\alpha_{1950} = 19^{h} 42^{m} 04^{s}.2$, $\delta_{1950} = -14^{\degr} 56^{'} 24^{''}$;][]{Gottes77} 
is such that it can be observed from both the northern and the southern hemisphere.
b) The true distance modulus based on the Tip of the Red Giant Branch (TRGB) is 
$\mu$ = 23.54 $\pm$ 0.05 mag \citep*{Lee93, Fusco12} that agree quite 
well with similar estimates based on different standard candles 
(classical Cepheids: \citealt{Gieren06, Madore09, Feast12, Rich14}; Mira variables: \citealt{White13}; carbon stars: \citealt{Parada21}). 
The reddening shows a clumpy distribution 
and the current estimates show a broad variety when moving from the region close 
to the galaxy center (E(B-V) $\sim$ 0.45 mag) to the outermost (E(B-V) $\sim$ 0.25 mag) 
regions \citep*{Massey95, Gallart96, Gieren06, Cannon06, Fusco12}. This means that 
NGC~6822 is a factor of two closer in distance and less affected by differential 
reddening when compared with IC~10. 
c) The metallicity distribution based on medium resolution spectra shows a
well defined peak located at [Fe/H]= -1.05 $\pm$ 0.01 and a standard deviation of 
$\sigma$= 0.49 dex \citep{Kirby13}. This measurement suggests that its mean 
chemical composition is, within the errors, quite similar to 
IC~1613 ([Fe/H]= -1.19 $\pm$ 0.01, $\sigma$ = 0.37 dex).  
However, this iron abundance refers to the old stellar population in NGC~6822.
Metallicity measurements for young stellar population have also been widely
investigated. From the spectra of two A-type supergiants
\citet{Venn01} found a mean iron abundance of [Fe/H]= -0.49 $\pm$ 0.22, while 
\citet{Lee06} obtained a mean oxygen abundance of [O/H]= -0.55 from optical 
spectra of five HII regions. More recently, \citet{Patrick15} found a mean 
metallicity of [Z]= -0.52 $\pm$ 0.21 by using medium-resolution near-IR spectra 
of eleven red supergiants (RSGs).
%_______________________________________________________________________________________
\begin{deluxetable}{lcccccc}
\tablenum{1}
\tablecaption{Positional, structural and photometric properties of NGC~6822.\label{tab:properties}}
\tablewidth{0pt}
\tablehead{
\colhead{Parameter} & & & & & & \colhead{Ref.$^{a}$}
} %\decimalcolnumbers
\startdata
$\alpha$ (J1950)$^{b}$   &\ &\ & $19^{h} 42^{m} 04^{s}.2$      &\ &\ & [1] \\
$\delta$ (J1950)$^{b}$   &\ &\ & $-14^{\degr} 56^{'} 24^{''}$  &\ &\ & [1] \\
$R_{half}$ ($'$)$^{c}$   &\ &\ & 16.13 $\pm$ 0.5$^{n}$         &\ &\ & [2] \\
$r_{t}$ ($'$)$^{d}$      &\ &\ & 40 $\pm$ 10			       &\ &\ & [3] \\
$\theta$ ($\degr$)$^{e}$ &\ &\ & 62					           &\ &\ & [2] \\
$\epsilon$ $^{f}$        &\ &\ & 0.35					       &\ &\ & [2] \\
$V$ (mag)$^{g}$          &\ &\ & 8.1 $\pm$ 0.2  		       &\ &\ & [4] \\
$M_V$ (mag)$^{h}$        &\ &\ & -15.2 $\pm$ 0.2		       &\ &\ & [5] \\
$(m-M)_{0}$ (mag)$^{i}$  &\ &\ & 23.38 $\pm$ 0.04		       &\ &\ & [6] \\
$E(B-V)$ (mag)$^{j}$     &\ &\ &  0.35 $\pm$ 0.04		       &\ &\ & [6] \\
$[Fe/H]^{k}$             &\ &\ & -1.05 $\pm$ 0.01 	           &\ &\ & [7] \\
$[Z]^{l}$                &\ &\ & -0.52 $\pm$ 0.21 	           &\ &\ & [8] \\
$[O/H]^{m}$              &\ &\ & -0.55           	           &\ &\ & [9] \\
\enddata
\tablecomments{$^{a}$ [1] \citet{Gottes77}; [2] \citet*{Zhang21}; [3] \citet{Hodge91}; [4] \citet{Dale07}; [5] \citet{McConnachie12} [6] \citet{Rich14}; [7] \citet{Kirby13}; [8] \citet{Patrick15}; [9] \citet{Lee06}.\\ 
$^{b}$ Galaxy center based on neutral hydrogen radio measurements. 
$^{c}$ Half-light radius. 
$^{d}$ Truncation radius. 
$^{e}$ Position angle. 
$^{f}$ Eccentricity.
$^{g}$ Apparent visual magnitude.
$^{h}$ Absolute visual magnitude.
$^{i}$ True distance modulus.
$^{j}$ Color excess.
$^{k}$ Mean iron abundance of the old population.
$^{l}$ Mean metallicity of the young population.
$^{m}$ Mean oxygen abundance of HII regions.
$^{n}$ The $R_{half}$ value was obtained by using the relation $R_{half}= 1.68r_0$ \citep*{Martin08} under the assumption of an exponential density profile and with the scale length provided by \citet[$r_0=9.6\pm0.3'$]{Zhang21}.}
\end{deluxetable}
%_______________________________________________________________________________________
%
The difference in metallicity between old and young 
stellar populations and the possible presence of a metallicity gradient in the 
innermost galaxy regions were explained by \citet{Patrick15} using a closed-box 
chemical evolution model. This model is based on plane physical arguments and takes 
account for the increase in the mean metallicity when moving from old to young 
stellar populations. However, NGC~6822 has a complex morphology, larger samples of 
old and young stellar tracers and more detailed chemical evolution models are 
required to constrain its chemical enrichment history.
The measurements of the $\alpha$-element abundances are quite limited and restricted 
to two Globular Clusters (GCs) of NGC~6822: \citet{Larsen18} have provided abundances 
for NGC~6822 SC6 ([Mg/Fe]= +0.295 with rms=0.240; [Ca/Fe]= +0.228 with rms=0.170; [Ti/Fe]= +0.290 with rms=0.091) 
and for NGC~6822 SC7 ([Mg/Fe]= -0.180 with rms=0.208; [Ca/Fe]= +0.042 with rms=0.141; [Ti/Fe]= +0.013 with rms=0.128). 
They found measurements consistent with those of nearby dwarf galaxies and also the first 
evidence of the knee in the [$\alpha$/Fe] versus [Fe/H] relation. 
d) NGC~6822 thanks to its modest distance has been the cross road of multi-band 
investigations ranging from the optical to the near-infrared (NIR) and to the 
mid-infrared (MIR) of Young Stellar Objects (YSO) and Asymptotic Giant Branch 
(AGB) stars (\citealt{Letarte02}, hereinafter~\citetalias{Letarte02}; \citealp*{Kinson21}; \citealp{Parada21}). These investigation have been soundly 
complemented by low- \citep*[hereinafter~\citetalias{Kacharov12}]{Kacharov12} and medium-resolution 
spectroscopy \citep[hereinafter~\citetalias{Sibbons12, Sibbons15}]{Sibbons12, Sibbons15}.
e) This galaxy has also aroused great interest
concerning the search, the identification and the characterization of variable 
stars. Optical time series data covering a time interval of months were collected by \citet{Pietr04}. 
They identified 116 classical Cepheids, together with sizable samples of Long Period Variables (LPVs) and eclipsing binaries \citep{Menni06}. 
More recently, a detailed census of LPVs and candidate LPVs were also provided by collecting NIR time series: \citet*{Battinelli11}, using the wide-field imager CPAPIR at the CTIO, have found 64 LPVs; \citet{White13}, using the IRSF at SIRIUS, have provided a list of 157 LPV stars. 
These investigation were soundly complemented with fainter time series data collected with FORS2 at VLT by \citet{Clementini03} and by \citet{Baldacci05}. They identified 
large samples of LPVs, eclipsing binaries, Chepheids and for the first time 
a sample of 41 candidate RR Lyrae (RRL) variables.   
f) NGC~6822 is one of the few nearby dwarf galaxies to host a spatially extended system 
of GCs. To date, have already been identified eight GCs belonging to the galaxy 
\citep{Hubble, Hwang11, Huxor13} and their structural properties are fundamental to 
probe the galaxy assembly process. 
g) NGC~6822 was one of the very first dwarf galaxies to be discovered and for this 
reason it held an important role in defining the scale of cosmic distances. 
The first distance estimate of NGC~6822 was determined by \citet{Shapley} and 
subsequently redefined by \citet{Hubble}. Its stellar content was a stepping stone 
for the concept of stellar populations and for the cosmic distance scale.
h) The mean metal-intermediate chemical composition and the identification of solid young 
(classical Cepheids), intermediate-age (LPVs) and old (RRLs) stellar tracers and its 
apparent isolation make this stellar system a very interesting laboratory for stellar 
pulsation and evolution and to investigate galactic evolution. 

The photometric investigation available in the literature and focused on NGC~6822 have 
moved along two different paths. 
a) {\em Small and deep} -- Detailed photometric investigations have been performed by 
using deep and accurate photometric images of several ACS at HST fields located across the 
disk of the galaxy by \citet{Cannon12} and by \citet{Fusco12, Fusco14}. 
The color-magnitude diagrams approached the turn off of the old stellar populations 
and the star formation histories based on these data provided detailed information 
of the radial distribution of young, intermediate and old stellar populations.  
The same outcome applies to the ground-based optical photometry collected with 
8m telescopes and to the space-based MIR photometry collected with SPITZER, 
because the field-of-view of the adopted detectors was quite limited and of 
the order of tens of arcmin squared.  
b) {\em Large and shallow} -- Several optical and NIR ground-based photometric 
investigations have also been performed by using wide field imagers at the 2-4m class 
telescopes. This means that they mainly focused either on young (YSOs, massive main sequence stars) or on intermediate-age (AGB) stellar tracers over the entire body of the galaxy.    

This is the first paper of a series aimed at studying the stellar content 
of NGC~6822 by taking advantage of optical photometry collected with wide field 
imagers available at the 4-8m class telescopes. In this investigation we will 
focus our attention on the photometric properties and the structure parameters 
of the galaxy, and discuss in detail the radial distribution of a broad range 
of stellar populations and the C/M ratio derived from the AGB stars count.

%%%%%%%%%%%%%%%%%%%%%%%%%%%%%%%%%%%%%%%%%%%%%%%%%%%%%%%%%%%%%%%%%%%%%%%%%%%%%%%%%%%%%
\section{Observations and data reduction} \label{sec:data}
%%%%%%%%%%%%%%%%%%%%%%%%%%%%%%%%%%%%%%%%%%%%%%%%%%%%%%%%%%%%%%%%%%%%%%%%%%%%%%%%%%%%%

%_______________________________________________________________________________________
\begin{deluxetable*}{cccccccc}
\tablenum{2}
\tablecaption{Summary of the observing runs for NGC~6822\label{tab:runs}}
\tablewidth{0pt}
\tablehead{
\colhead{Run name} & \colhead{Dates} & \colhead{Telescope} & \colhead{Camera} &
\colhead{N$_{r}$} & \colhead{N$_{g}$} & \colhead{N$_{i}$} & \colhead{multiplex} \\
} %\decimalcolnumbers
\startdata
suba29  & 2014 Sep 29        & Subaru 8.2m & Hyper Suprime-Cam & 18 & -- & -- & x 104 \\
suba29  & 2015 Oct 07        & Subaru 8.2m & Hyper Suprime-Cam & -- & 25 & 12 & x 104 \\
suba29  & 2016 Jun 07        & Subaru 8.2m & Hyper Suprime-Cam & -- & 36 & 15 & x 104 \\
mp6822  & 2004 May 26-Jun 25 & CFHT 3.6m   & MegaPrime         & 3  & 3  & 7  & x 36  \\
dec1605 & 2016 May 28        & CTIO 4m     & DECam             & 4  & 4  & 4  & x 34  \\
dec1607 & 2016 Jul 04-05     & CTIO 4m     & DECam             & 4  & 4  & 4  & x 34  \\
int0305 & 2003 May 01-03     & INT 2.5m    & Wide Field Camera & -- & 6  & -- & x 4   \\
\enddata
\tablecomments{The most external HSC CCDs have not been used to obtain our photometric catalog.}
\end{deluxetable*}
%_______________________________________________________________________________________

Our photometric catalog was obtained by using a sizable sample of multi-band ($g,r,i$) images 
collected with Hyper-Suprime-Cam (HSC) at Maunakea Subaru Telescope. 
The individual images were collected by rotating the camera of 45 degrees 
in consecutive exposures to fill the gaps among CCDs, and to overcome difficulties 
with bright saturated stars (blooming) and with artifacts on individual CCDs.
%_______________________________________________________________________________________
\begin{figure}[htbp!]
\centering
\includegraphics[width=8cm]{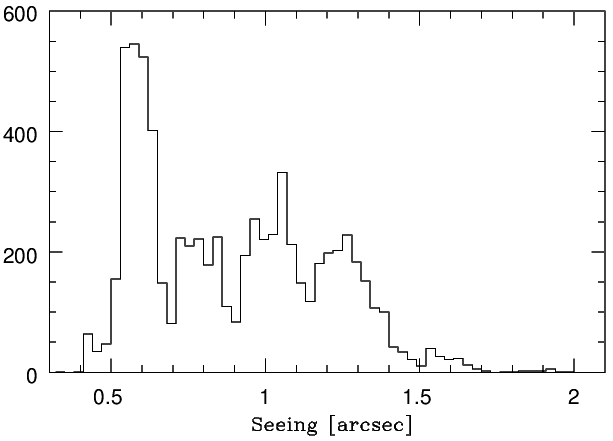}
\caption{Seeing distribution of individual CCD images included in the dataset. \label{fig:seeing}}
\end{figure}
%_______________________________________________________________________________________
%
The HSC has a field of view (FoV) of 1.5 degrees in diameter, with a 0.17$''$/pixel scale. 
These images include both deep and shallow exposures, acquired during three different 
nights: September 29th, 2014 ($r$-band); October 7th, 2015  and June 7th, 2016 ($g$ and $i$-band). 
The shallow exposures include 4~$r$, 1~$g$ and 3~$i$-band images, each with an exposure time of 
30~$s$. The deep exposures include 14~$r$, 60~$g$ and 24~$i$-band images, with exposure times 
of 300~$s$ for the $r$-band and 240 $s$ for both the $g$ and the $i$-band.
The HSC is equipped with 104 CCDs, but the current photometric catalog does not include a few 
of the most external ones.

%_______________________________________________________________________________________
\begin{figure*}
\gridline{\fig{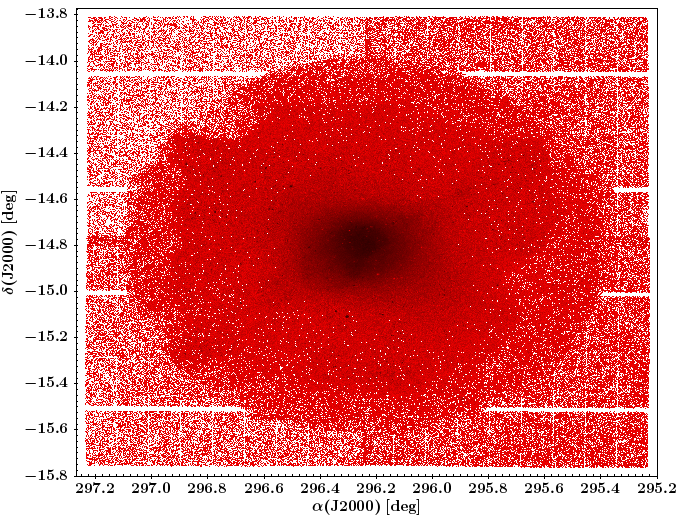}{0.52\textwidth}{}
          \fig{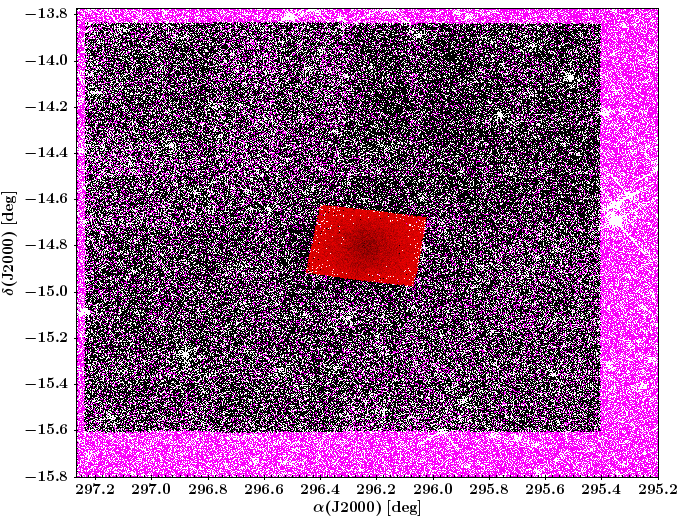}{0.52\textwidth}{}
          }
\caption{Left: sky coverage of the HSC (shallow and deep), MegaPrime, DECam and WFC $g$, $r$ and $i$ images.
Right: same as the left panel but for the WISE (magenta dots), UKIRT (black dots) and Spitzer 
(red dots) datasets.  \label{fig:maps}}
\end{figure*}
%_______________________________________________________________________________________

This dataset was complemented with multi-band images collected with the wide field imager 
MegaPrime available at the Canada-France-Hawaii Telescope (CFHT), with the Dark Energy 
Camera (DECam) on the Blanco telescope at Cerro Tololo Inter-American Observatory (CTIO), 
and with the Wide Field Camera (WFC) mounted on the Isaac Newton Telescope (INT) at La Palma.
Actually, these three optical datasets were included to improve 
the absolute photometric calibration and the sampling of bright stars. 
The images from the CFHT were acquired between May 26th and June 25th, 2004 and 
include 3~$r$, 3~$g$ and 7~$i$-band images and their exposure times range from 15 to 150 $s$. 
The MegaPrime has a FoV of 1x1 square degree with a 0.19$''$/pixel scale. 
The data acquired at the Blanco telescope include 8~$r$, 8~$g$ and 8~$i$-band images 
collected over three nights (May 28th, July 4th and 5th, 2016) and their exposure times range from 
30 to 80 $s$. The DECam has a FoV of 2.2 square degrees in diameter with a 0.27$''$/pixel scale.
Finally, the images from the INT include 6~$g$-band exposures collected during two nights on 
1st and 3rd May, 2003 and have an exposure time of about 600 $s$. The WFC has a 
34$\times$34 arcmin FoV with a 0.33$''$/pixel scale.
All in all the current dataset cover 2 square degrees across the center of the galaxy and a time 
interval of 13 years. The seeing distribution of individual CCD images included in the dataset 
is plotted in Fig.~\ref{fig:seeing}, it suggests that 25\% of the images were collected with a
seeing better than 0.6$''$ and 60\% with a seeing better than 1$''$.
The log of the observing runs is summarized in Table \ref{tab:runs}.
Fig.~\ref{fig:maps} shows the sky coverage of the different datasets used in this study. The left panel displays the coverage of the HSC (shallow and deep), the MegaPrime, the DECam and the WFC $g$, $r$ and $i$-band exposures. The right panel shows the sky coverage of the infrared (IR) catalogs available in literature for NGC~6822\footnote{The NIR (\citetalias{Sibbons12}) and MIR \citep{Khan15} catalogs are available on the VizieR database at \url{https://vizier.u-strasbg.fr/viz-bin/VizieR}}.
i) Near-Infrared (NIR) catalog (black dots; \citetalias{Sibbons12}) covering an area of 1.8 square degrees across the galaxy center and acquired at the 3.8 m United Kingdom Infrared Telescope (UKIRT) with the Wide Field CAMera (WFCAM; 0.4$''$/pixel scale) as part of the UKIRT Infrared Deep Sky Survey 
(UKIDSS, \citealt{Warren07}).
ii) Mid-Infrared (MIR) photometry (red dots; \citealt{Khan15}) covering the innermost 
(5.2$\times$5.2~arcmin squared) central regions and collected from the mosaics of the 
SIRTF Nearby Galaxies Survey \citep[SINGS;][]{Kenni03} by using the Infrared Array Camera 
(IRAC; 1.2$''$/pixel scale) on board Spitzer Space Telescope (Spitzer). 
iii) MIR photometry (magenta dots) obtained with the Wide-field
Infrared Survey Explorer (WISE; \citealt{Wright10}), covering an area of
3$\times$3~degrees centered on the galaxy. The data have been extracted
from the CatWISE2020 catalog \citep{Marocco21} at 3.4 and 4.6~$\mu m$
(W1 and W2 band, respectively), and from the AllWISE release of the WISE
Source Catalog \citep{Cutri13} at 12~$\mu m$ (W3 band).
The reasons for which we decided to use the NIR and MIR catalogs 
will become more clear in the following sections.

The data reduction was mainly performed with the DAOPHOT-ALLSTAR-ALLFRAME packages \citep{Stetson87, Stetson94}. For the HSC images we adopted the following procedure: as a first step, the point-spread functions (PSFs) of the stars on individual images have been computed through a custom automatic pipeline, based on a suite of DAOPHOT/ALLSTAR SExtractor \citep{Bertin96} Python Fortran routines, which discards outliers, extended sources, blends and performs the WCS alignment. Subsequently, the ALLFRAME routine was used to improve the PSF photometry on the different sets of CCDs covering the same FoV area. The instrumental magnitudes thus obtained were calibrated to the Pan-STARRS $g,r,i$ standard system by using a list of 41,720 stars from the Pan-STARRS catalog. The derived transformation equations for each image include first- and second-order colour terms, zero-points and linear positional terms (X and Y), and have standard deviations of 0.015, 0.012 and 0.010 mag in the $g$, $r$ and $i$-band, respectively. 
%_______________________________________________________________________________________
\begin{figure}[ht!]
\centering
\includegraphics[width=8.5cm]{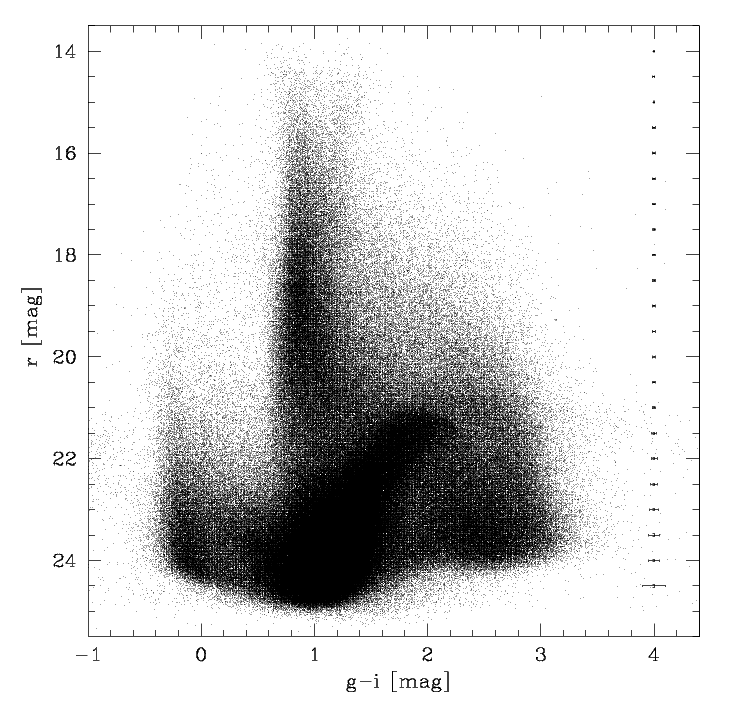}
\caption{\textit{r}, \textit{g-i} CMD based on the entire optical dataset. The error 
bars plotted on the right side of the CMD (\textit{g-i} $\sim$ 4 mag) display the 
intrinsic errors both in magnitude and in color (summed in quadrature). \label{fig:error}}
\end{figure}
%_______________________________________________________________________________________
%
The final combined and photometrically calibrated catalog contains more than 1 million objects 
and the limiting magnitudes range from 14 to 25.8 mag in the $g$-band, the deepest, 
from 13.7 to 25.5 mag in the $r$-band, and from 13 to 24.9 mag in the $i$-band, the shallowest,
thus providing the widest and most homogeneous photometric dataset ever collected 
for a nearby dIrr, but the Magellanic Clouds. 
The photometric errors typically range from 0.003 up to 0.01 mag for $g,r,i\leq$ 20 mag; from 0.01 up to 0.02 mag for 20 $<g,r,i\leq$ 22 mag; and increase quite steeply from 0.02 up to 0.09 for $g,r,i>$ 22 mag.
The \textit{r}, \textit{g-i} color-magnitude diagram (CMD) resulting from the entire optical dataset is plotted in Fig.~\ref{fig:error}, together with the error bars showing the intrinsic photometric errors both in magnitude and in color, for which they were summed in quadrature. The CMD shows some clear features of the NGC~6822 galaxy, such as the Young Main Sequence (YMS) at $g-i\sim -0.2-0$ mag and the Red Giant Branch (RGB) ranging over about 3.5 mag in magnitudes and 1.2 mag in color, but also a considerable number of foreground and background field stars, indeed the two sequences of G-type stars at $g-i\sim 1$ mag and M-type stars at $g-i\sim 2-3$ mag are particularly prominent.

%%%%%%%%%%%%%%%%%%%%%%%%%%%%%%%%%%%%%%%%%%%%%%%%%%%%%%%%%%%%%%%%%%%%%%%%%%%%%%%%%%%%%
\section{Identification of candidate field and galaxy stars} \label{sec:separation}
%%%%%%%%%%%%%%%%%%%%%%%%%%%%%%%%%%%%%%%%%%%%%%%%%%%%%%%%%%%%%%%%%%%%%%%%%%%%%%%%%%%%%

The separation between candidate field and galaxy stars is one of the fundamental 
issues concerning the analysis of stellar populations in nearby stellar systems. 
Stellar systems located at low Galactic latitudes are severely affected by field 
contamination, therefore, star counts and the comparison between theory and observations 
in both optical and near-infrared CMDs are hampered by foreground and background 
Galactic stars. The problem becomes even more severe for gas rich stellar systems, 
because they are typically affected by differential reddening. NGC~6822 is affected 
by the quoted limitations and to overcome these difficulties special attention was 
paid to the separation between field and galaxy stars. 

%_______________________________________________________________________________________
\begin{figure*}[ht!]
\gridline{\fig{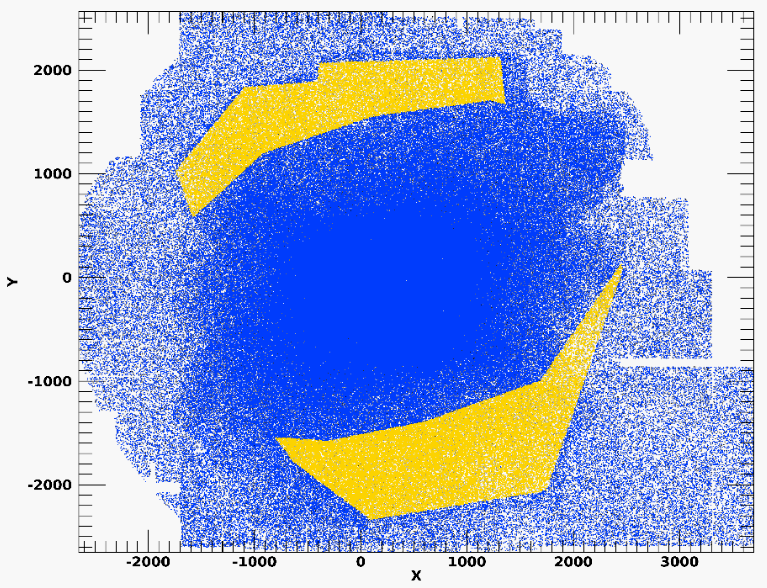}{0.43\textwidth}{(a)}
          \fig{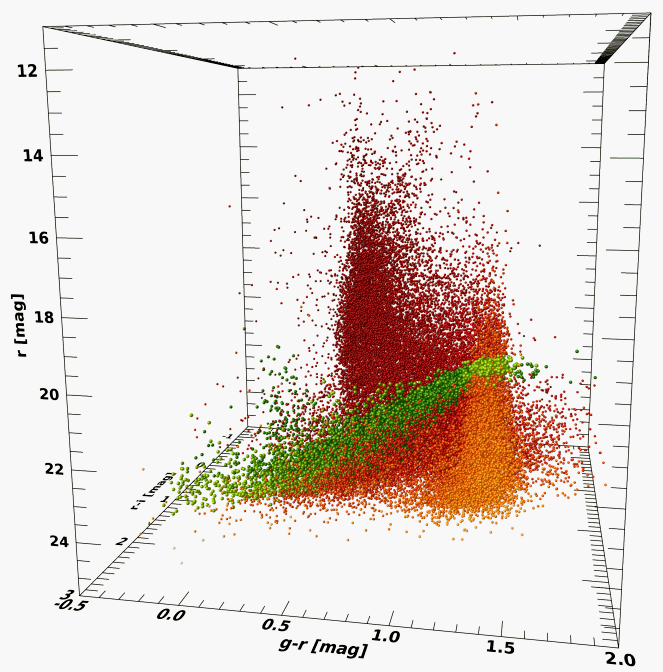}{0.35\textwidth}{(b)}
          }
\gridline{\fig{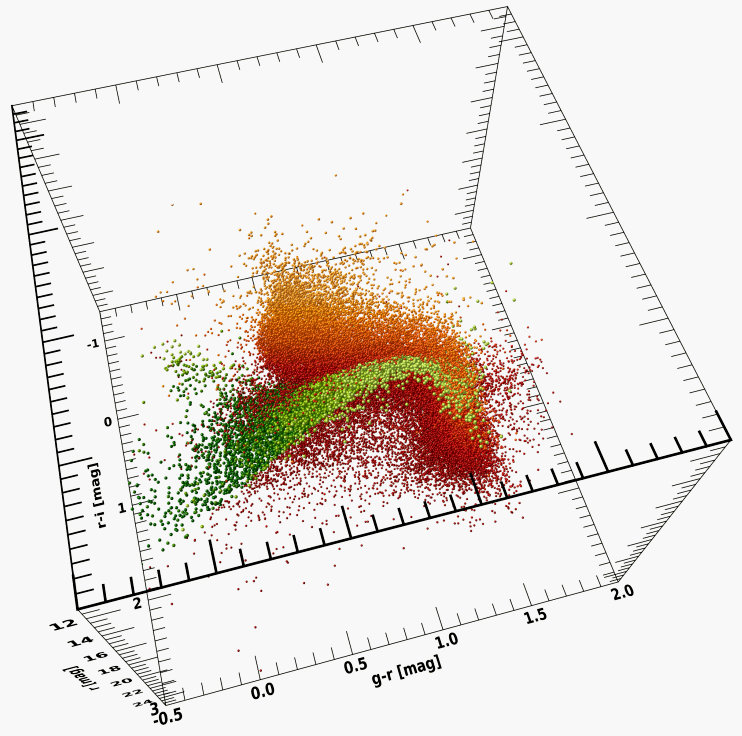}{0.35\textwidth}{(c)}
          \fig{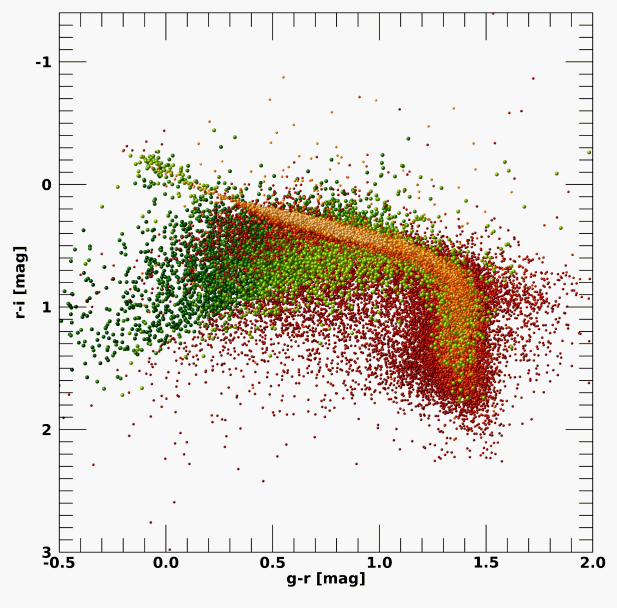}{0.35\textwidth}{(d)}
          }
\gridline{\fig{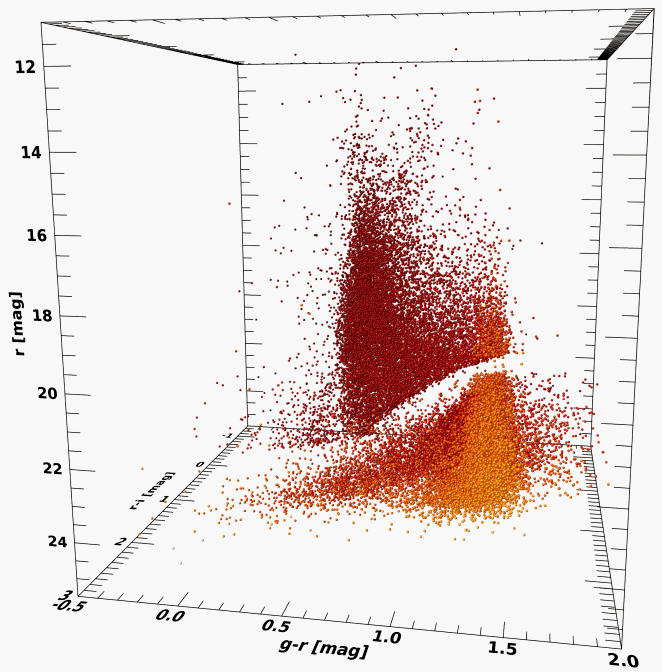}{0.35\textwidth}{(e)}
          \fig{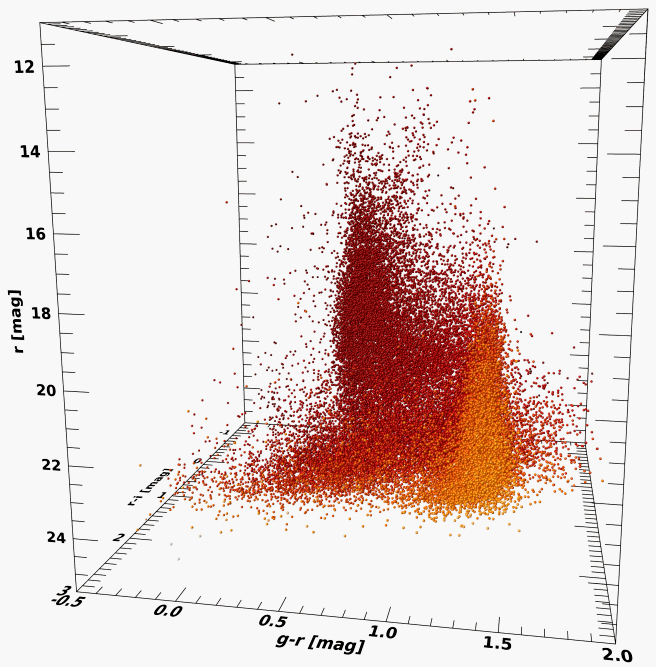}{0.35\textwidth}{(f)}
          }          
\caption{
Panel a)-- Distribution on sky of the entire optical catalog (blue dots). The yellow dots 
display the regions adopted for selecting candidate galaxy stars. 
Panel b)--  \textit{g-r}, \textit{r-i}, \textit{r} Color-Color-Magnitude diagram (CCMD). 
Green dots display candidate galaxy stars, while red dots candidate field stars.   
Panel c)-- Same as Panel b), but with a different view angle.   
Panel d)-- Same as Panel b), but the view is from the top of the z-axis, i.e. projected 
onto the  \textit{g-r}, \textit{r-i}, color-color diagram (CCD).  
Panel e)-- Same as Panel b), but we only plotted candidate field stars after the subtraction 
of candidate galaxy stars.  
Panel f)-- Same as Panel e), but the gap was filled by statistically subtracting stars from 
the sample of candidate galaxy stars. See text for more details.    
\label{fig:separationGC}}
\end{figure*}
%_______________________________________________________________________________________
%
%_______________________________________________________________________________________
\begin{figure*}[htbp!]
\centering
\includegraphics[width=15cm]{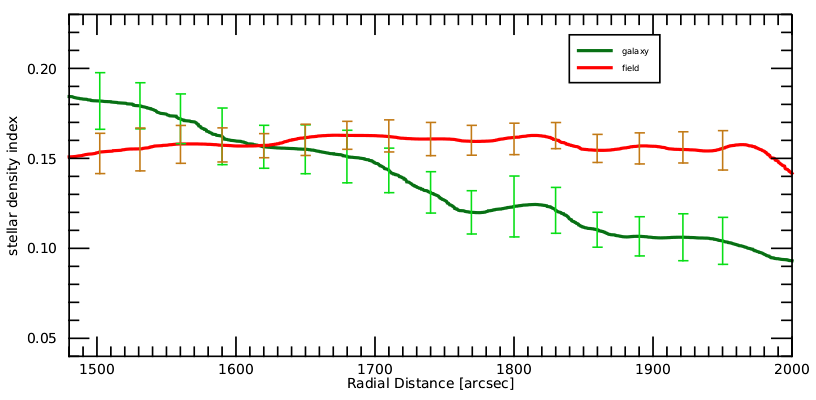}
\caption{Surface density of candidate galaxy (green line) and field (red line)
stars as a function of the radial distance for the stars marked with
yellow dots in Fig.~\ref{fig:separationGC}a. The radial distributions were
smoothed by using a Gaussian kernel with unitary weight. The error
bars display the standard deviation in the estimate of the surface
density. \label{fig:dens4}}
\end{figure*}
%_______________________________________________________________________________________

The approach we devised differs from the approach we have already adopted for 
$\omega$ Cen \citep{Calamida20} and nearby dwarf galaxies
Carina \citep{Bono10} and Sculptor \citep{Martinez16}. 
NGC~6822 includes young, intermediate and old stellar tracers, therefore, we developed 
a new method  based on 3D color-color-magnitude diagrams (CCMDs).
The initial step relies on the separation between the most relevant evolutionary
features of candidate field and galaxy stars. To accomplish this goal we selected
two different regions located well outside the central regions of the galaxy
(see yellow dots plotted in Fig.~\ref{fig:separationGC}a). The reasons for 
this selection is twofold:

a) These regions are located outside the disk/bar and beyond the core radius.
This means that they are less affected by crowding, therefore, their
photometry is deeper and more accurate. Moreover, these regions are also less
affected by differential reddening and the evolutionary sequences are intrinsically
narrower.

b) Star counts of both field and galaxy stars almost balance at these radial distances.
This means a proper identification of the key evolutionary features. Indeed, at
larger radial distances the CMD is dominated by field stars, whereas at smaller
radial distances by galaxy stars. Note that the sharp decrease in the stellar
density of the most external regions is a consequence of the fact that these
regions were only covered with shallow exposure times. Therefore, their limiting
magnitude is 2/3 mag shallower than for the internal regions.

Fig.~\ref{fig:separationGC}b shows the 3D CCMD, $g-r$,$r-i$,$r$.
We performed a preliminary identification of both candidate field stars, plotted
with orange/red dots, and candidate galaxy stars plotted on top with green dots.
The different sequences were identified by using different view angles that help
in the identification of both field and galaxy sequences. Moreover, to properly 
identify candidate galaxy stars we also used different sets of stellar isochrones 
transformed into the PanSTARRS photometric system from the BASTI 
database\footnote{The interested reader is referred to 
\url{http://basti-iac.oa-abruzzo.inaf.it/index.html}.}
covering a broad range in stellar ages and chemical compositions (see discussion 
in Section~\ref{sec:tracers}). To make more clear the adopted procedure, 
Fig.~\ref{fig:separationGC}c shows the same stars in the CCMD, but the
view is from the top of the Z-axis ($r$ mag). Data plotted in these CCMDs
allow us to properly identify candidate field stars, namely the sequence of
G-type stars located at $g-r\sim$ 0.4--0.8 mag and the more relevant sequence of
M-type stars located at $g-r\sim$ 1.2--1.5 mag \citep{Castellani02, Girardi12, Robin14}.
After a number of test and trial in which we took into account the luminosity and the 
color function we also performed a preliminary identification of galactic RGB and 
early AGB sequences (green dots). To further highlight the difficulties in 
disentangling galaxy and field sequences, Fig.~\ref{fig:separationGC}d shows 
the same stars, but plotted in the $g-r$,$r-i$ color-color Diagram (CCD).

Note that candidate galaxy and field stars overlap over the entire color ranges.
A glance at the data plotted in this CCD clearly shows that along the line of
sight and at the quoted radial distances, field stars outnumber candidate galaxy
stars. Selections only based on colors will include a significant fraction
of candidate field stars among candidate galaxy stars and viceversa.
As a final validation of the selection criteria adopted to identify candidate 
field and galaxy stars we also investigated their radial distributions. 
Fig.~\ref{fig:dens4} show the radial distribution of the stars marked with 
yellow dots in Fig.~\ref{fig:separationGC}a. 
We estimated the surface density of candidate field (red line) and galaxy 
(green line) stars in concentric annuli of equal area and plotted them 
as a function of the radial distance. The radial distributions were also 
smoothed by assuming a Gaussian kernel with unitary weight. Data plotted in this 
figure show that candidate field stars have, within the errors, a flat distribution 
over the entire area, while the candidate galaxy stars display, as expected, a steady 
decrease when moving from the innermost to the outermost galaxy regions. 

Subsequently, we removed the sequence of candidate galaxy stars and were 
left with the CCMD showed in Fig. \ref{fig:separationGC}e. Data plotted in this figure 
show that the subtraction of candidate galaxy stars caused an over 
subtraction of field stars in the region with colors ranging 
from $g-r \sim$ 0.0 to $g-r \sim$ 1.5 and from $r-i$ $\sim$ 0.7 mag to $r-i$ $\sim$ 1.1 mag,
and with $r$ magnitudes ranging from $r\sim$ 23 to $r\sim$ 18.5 mag. 
To overcome this problem we split the same 
CCMD region of the candidate galaxy stars (see green dots in Fig. \ref{fig:separationGC}b) 
in roughly two dozen of contiguous slabs. We randomly extracted in each slab 
of candidate galaxy stars a number of stars that allowed us to have in 
each slab of candidate field stars a number of stars similar (within Poisson 
uncertainties) to the regions of the CCMD located at the edges of the gap 
caused by the subtraction of candidate galaxy stars. The outcome of the 
random extraction of possible field stars among candidate galaxy stars 
is displayed in Fig. \ref{fig:separationGC}f, in which the gap is no more present.   
The new catalog of candidate field stars plotted in Fig.~\ref{fig:separationGC}f
was statistically subtracted to the entire photometric catalog (yellow
plus blue dots plotted in Fig.~\ref{fig:separationGC}a). The main outcomes of 
this process were two new subsamples of candidate field and galaxy stars. 
The latter subsample was used to improve the selection criteria adopted to 
separate field and galaxy stars (yellow dots in Fig.~\ref{fig:separationGC}a) 
and following once again the same steps discussed at the beginning of this section we 
ended up with a final catalog of candidate galaxy stars including more than 
550,000 stars with at least one measurement in two different photometric bands.

%_______________________________________________________________________________________
\begin{figure*}[ht!]
\centering
\includegraphics[width=15cm]{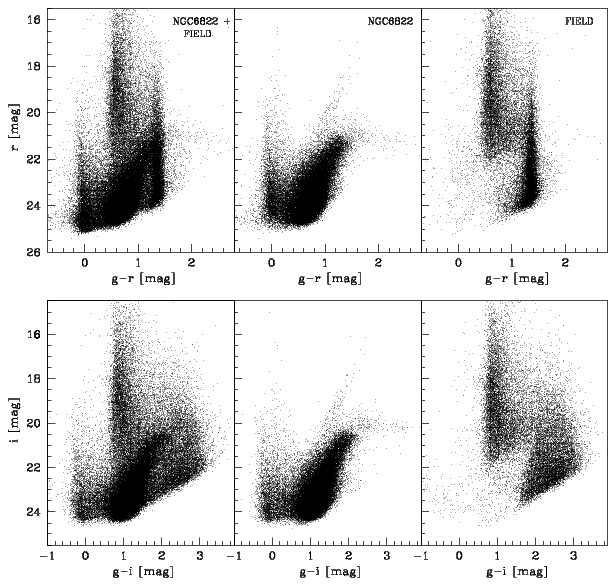}
\caption{Left: \textit{r}, \textit{g-r} (upper panel) and \textit{i}, \textit{g-i} (lower panel) CMDs of the entire optical catalog. 
Middle: same as the left panels, but for candidate NGC~6822 stars. The data in this panel display several well defined evolutionary phases: Young MS stars (16 $\le$ \textit{i} $\le$ 24.5 mag, \textit{g-i} $\sim$ 0  mag), RSGs (16 $\le$ \textit{i} $\le$ 20.5 mag, 1.4 $\le$ \textit{g-i} $\le$ 2 mag), the tip of the RGB (\textit{i} $\sim$ 20.5 mag, \textit{g-i} $\sim$ 2 mag), and AGB stars (19.5 $\le$ \textit{i} $\le$ 20.5 mag, 1.8 $\le$ \textit{g-i} $\le$ 3.8 mag). 
Right: same as the left panels, but for candidate field stars. 
The CMDs showed in this figure only display one star every ten of the global catalog. 
\label{fig:gal+field}}
\end{figure*}
%_______________________________________________________________________________

The final outcome is shown in Fig. \ref{fig:gal+field}. The top panel displays the  \textit{r}, \textit{g-r} CMDs for the whole photometric catalog (left), stars candidate to belong to NGC~6822 (middle) and field stars (right); the bottom panel is the same as the top panel, but for \textit{i}, \textit{g-i} CMDs. 
For the sake of clearness, we decided to plot one object over ten selected from each catalogs.
Note that, when field stars were removed, some well defined evolutionary phases can be easily recognized: 
Young MS stars (16$\le i \le$24.5 mag, $g-i \sim$ 0  mag), 
RSGs (16$\le i \le$20.5 mag, 1.4$\le g-i \le$2 mag),    
the tip of the RGB ($i \sim$ 20.5 mag, $g-i \sim$ 2 mag), 
and AGB stars (19.5$\le i \le$20.5 mag, 1.8$\le g-i \le$ 3.8 mag). 
The candidate field stars plotted in the right panels show that the separation between field and galaxy stars can be improved, indeed, in the region between 19.5$\le i \le$22.5 mag and 1.2$\le g-i \le$1.8 mag, some galaxy stars could also be present.

%_______________________________________________________________________________________
\begin{deluxetable*}{cccccc}
\tablenum{3}
\tablecaption{Total apparent and absolute magnitudes for NGC~6822, the SMC and the LMC.\label{tab:mag}}
\tablewidth{0pt}
\tablehead{
\colhead{Band} & \colhead{Apparent} & \colhead{Absolute} & \colhead{Absolute$^{a}$} & \colhead{Absolute for SMC$^{c}$} & \colhead{Absolute for LMC$^{c}$} \\
\colhead{}     & \colhead{(mag)}    & \colhead{(mag)}    & \colhead{(mag)}	   & \colhead{(mag)}           & \colhead{(mag)}
} %\decimalcolnumbers
\startdata
g       &\ 8.43 $\pm$ 0.03	 &\ \ \ -15.03 $\pm$ 0.07	&\ $\ldots$ &\ $\ldots$ 	 &\ $\ldots$	      \\
r       &\ 7.97 $\pm$ 0.03	 &\ \ \ -15.49 $\pm$ 0.07	&\ $\ldots$ &\ $\ldots$ 	 &\ $\ldots$	      \\
i       &\ 7.68 $\pm$ 0.04	 &\ \ \ -15.78 $\pm$ 0.08	&\ $\ldots$ &\ $\ldots$ 	 &\ $\ldots$	      \\
B       &\ 8.85 $\pm$ 0.05$^{b}$ &\ \ \ -14.61 $\pm$ 0.08$^{b}$ &\ -14.62   &\ -16.89 $\pm$ 0.06 &\ -18.28 $\pm$ 0.05 \\
V       &\ 8.93 $\pm$ 0.03$^{b}$ &\ \ \ -14.53 $\pm$ 0.07$^{b}$ &\ -15.09   &\ -16.29 $\pm$ 0.06 &\ -17.60 $\pm$ 0.05 \\
J       &\ 7.91 $\pm$ 0.10	 &\ \ \ -15.55 $\pm$ 0.12	&\ -17.35   &\ -17.35 $\pm$ 0.06 &\ -18.77 $\pm$ 0.05 \\
H       &\ 7.29 $\pm$ 0.08	 &\ \ \ -16.17 $\pm$ 0.10	&\ -17.83   &\ -18.05 $\pm$ 0.06 &\ -19.52 $\pm$ 0.05 \\
K       &\ 7.10 $\pm$ 0.10	 &\ \ \ -16.36 $\pm$ 0.12	&\ -18.29   &\ -18.27 $\pm$ 0.06 &\ -19.77 $\pm$ 0.05 \\
W1      &\ 5.86 $\pm$ 0.09	 &\ \ \ -17.60 $\pm$ 0.11	&\ -18.08   &\ -18.24 $\pm$ 0.06 &\ -19.60 $\pm$ 0.05 \\
W2      &\ 5.76 $\pm$ 0.14	 &\ \ \ -17.70 $\pm$ 0.15	&\ -18.25   &\ -18.26 $\pm$ 0.06 &\ -19.59 $\pm$ 0.05 \\
W3      &\ 4.12 $\pm$ 0.33	 &\ \ \ -19.34 $\pm$ 0.34	&\ $\ldots$ &\ -20.44 $\pm$ 0.06 &\ -21.27 $\pm$ 0.05 \\
\enddata
\tablecomments{$^{a}$ Absolute magnitudes from the literature: $V$ and $B$-band magnitudes were obtained by using the total apparent magnitudes from \citet{Mateo98} and the same reddening and true distance modulus used for our estimates; Total NIR/MIR magnitudes were estimated by using the global unreddened flux densities from \citet{Dale07} and the same true distance modulus used for our estimates. Note that we neglected uncertainties on individual estimates, since they are very similar to the current ones. 
$^{b}$ $V,B$ total apparent and absolute magnitudes were obtained by using the linear transformations between PanSTARRS and Johnson/Cousins photometric systems from \citet{Tonry12} (see text for more details).
$^{c}$ $V,B$ total absolute magnitudes were obtained by using the colour-colour transformations between Gaia eDR3 and Johnson/Cousins photometric systems from \citet{Riello21} (see text for more details). $J,H,K$ in the 2MASS system and $W1,W2,W3$ in the AllWISE system were estimated from measurements in the Gaia eDR3 catalogue.}
\end{deluxetable*}
%_______________________________________________________________________________________

The catalog of the candidate galaxy stars allows us to provide accurate estimates 
of the absolute magnitude of NGC~6822 in all of the investigated photometric bands.
The mean apparent and absolute magnitudes are listed in Table~\ref{tab:mag} together with their uncertainties. 
We unreddened the apparent magnitudes of candidate galaxy stars by using the 
reddening map provided by \citet*[$E(B-V) = 0.24$ mag]{Schlegel98} and the reddening law provided 
by \citet*{Cardelli89}. 
Note that we decided to neglect the intrinsic reddening, since according to 
the reddening map by \citet{Schlegel98} the variation across the main body of 
the galaxy is $\Delta E(B-V)=0.04$ mag. This means a minimal impact on 
the absolute magnitudes listed in Table~\ref{tab:mag}.
Then the apparent unreddened magnitudes were transformed 
in fluxes, summed up and transformed into magnitudes once again. Subsequently, the 
total apparent magnitudes were transformed into absolute magnitudes by adopting 
a true distance modulus of $\mu$ = 23.46 $\pm$ 0.06 mag, that is the mean value 
of the true distance moduli obtained by \citet{Fusco12}, by using the TRGB, 
and by \citet{Rich14} by using classical Cepheids. 
The $g,r$ total apparent and mean absolute magnitudes were transformed into $B,V$ 
apparent and absolute magnitudes by using the following linear transformations between 
PanSTARRS and Johnson/Cousins photometric systems provided by \citet{Tonry12}: 
$g-r=0.213 + 0.587(B-g)$ and $g-r=0.006 + 0.474(V-r)$.
More detailed transformations based on the $BVRI$ and on the $g,r,i$ photometry 
available to our group will be provided in a forthcoming paper.   
Note that in order to estimate the total apparent magnitudes in $W1$ and $W2$ bands
we took into account for the $[3.6]$ and $[4.5]$ measurements of candidate galaxy stars, 
covering the central regions and collected with IRAC@SPITZER, together with the $W1$ 
and $W2$ measurements covering the more external regions and collected by WISE. 
Actually, \citet{Cutri13} found that the W1 and W2 bands photometry is
consistent with the IRAC@SPITZER magnitudes in the two equivalent bands
$[3.6]$ and $[4.5]$, respectively, with only small differences largely due to the
small changes in bandpasses. This result has been confirmed in
subsequent studies that found a negligible difference between the W1 and
IRAC $[3.6]$ bands, and an offset of 0.028~mag between the W2 and IRAC
$[4.5]$ bands \citep{Papovich16}. This difference is reduced to
less than 0.02~mag for late spectral-type stars.
The column 4 of the same table gives the absolute magnitudes available 
in the literature for NGC~6822. The $B$, $W1$ and $W2$ magnitudes are quite similar, 
within the errors, to the current estimates. On the contrary, the $J,H,K$ absolute magnitudes 
are, on average, two magnitudes fainter. The difference is mainly due to the fact that
the $J,H,K$ literature estimates are based on 2MASS photometry that is shallower than 
UKIRT photometry and typically they did not perform a separation between candidate 
field and galaxy stars.
The columns 5 and 6 of the table list the total absolute magnitudes for the Large Magellanic Cloud (LMC)
and the Small Magellanic Cloud (SMC) in the same photometric bands used for NGC~6822, 
with the exception of $g,r,i$. 
Apparent magnitudes for indiviual stars in the MCs are available in the Gaia eDR3 
catalogue which provides measurements in the three Gaia photometric bands 
$G,G_{BP},G_{RP}$, in the 2MASS NIR ($J,H,K$) and in the AllWISE MIR ($W1,W2,W3$) 
bands (see Appendix~B for more details).  
The individual apparent magnitudes were unreddened by using the reddening values provided 
by \citet[$E(B-V)_{LMC}=0.091\pm0.050$ mag, $E(B-V)_{SMC}=0.038\pm0.053$ mag]{Joshi19}, 
and then the total absolute magnitudes were estimated by using the true distance moduli from 
\citet[$\mu_{LMC}= 18.477 \pm 0.004$ mag]{Pietr19} and 
\citet[$\mu_{SMC}= 18.977 \pm 0.016$ mag]{Graczyk20}.
Finally, the $G,G_{BP},G_{RP}$ total absolute magnitudes were transformed into $B,V$ absolute magnitudes
by using the following colour-colour transformations between Gaia eDR3 and Johnson/Cousins 
photometric systems from \citet{Riello21}: 
$G-V = -0.02704 + 0.01424(G_{BP}-G_{RP}) - 0.2156(G_{BP}-G_{RP})^2 + 0.01426(G_{BP}-G_{RP})^3$ and
$G-V = -0.04749 - 0.0124(B-V) - 0.2901(B-V)^2 + 0.02008(B-V)^3$.
The results show that NGC~6822 in all the listed photometric bands is fainter than the MCs. 
Moreover, the magnitudes in Table~\ref{tab:mag} indicate that NGC~6822 has optical-NIR colors
similar to those of the MCs, and optical-MIR colors that are systematically redder than MCs
with a difference of the order of one magnitude. This difference is mainly caused by 
the large relative number of C-rich stars identified in NGC~6822. 

In this context it is worth mentioning that the total apparent magnitudes in the three 
WISE photometric bands can be used as diagnostics to identify active galactic nuclei 
(AGNs) in dwarf galaxies. Indeed, \citet{Hain16} used the infrared WISE color-color 
($W1-W2$, $W2-W3$) diagnostic diagram to separate the AGNs and the so-called ``composite" 
galaxies, defined as stellar systems having contributions to their emission-line flux 
from both star-formation and AGN activity \citep*{Baldwin81, Kewley01, Kauffmann03}.
According to the current estimates, NGC~6822 has unreddened WISE colors of $W1-W2$ = 0.10 mag 
and $W2-W3$ = 1.64 mag. This means that NGC~6822 is located along the blue tail of the 
``composite" galaxy sequence (see Fig.~1 of \citet{Hain16}). 
This evidence taken at face value indicates that in the 
selected sample of composite galaxies the AGB stars, and in particular, C-rich stars 
can contribute to their total flux, together with young stars (YMS, RSG) and possibly 
with AGN activity. The current findings suggest that MIR color space should be 
cautiously treated in the selection of AGNs in dwarf galaxies due to the non-trivial 
contribution of young \citep{Reines22}, and in particular, of intermediate-age 
stellar population.

%%%%%%%%%%%%%%%%%%%%%%%%%%%%%%%%%%%%%%%%%%%%%%%%%%%%%%%%%%%%%%%%%%%%%%%%%%%%%%%%%%%%%
\section{A new estimate of the center of the galaxy} \label{sec:center}
%%%%%%%%%%%%%%%%%%%%%%%%%%%%%%%%%%%%%%%%%%%%%%%%%%%%%%%%%%%%%%%%%%%%%%%%%%%%%%%%%%%%%

%_______________________________________________________________________________
\begin{figure*}[ht!]
\centering
\includegraphics[width=18.8cm]{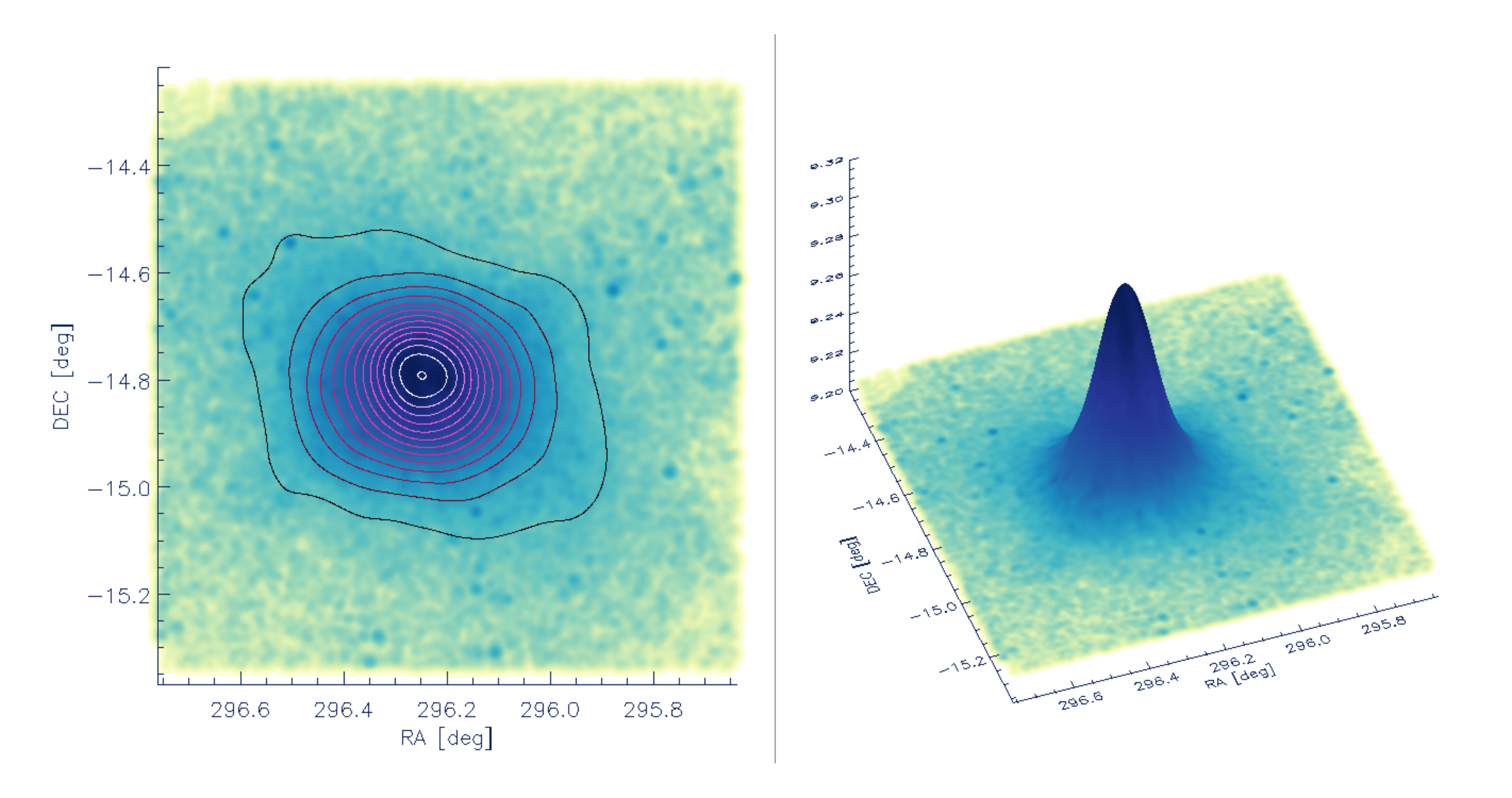}
\caption{Left: radial distribution on sky of the old stellar tracers (RGB$+$RHB), smoothed by assuming 
a Gaussian kernel with unitary weight. Overplotted are the isocontour levels. 
Right: 3D histogram (RA, DEC, N in arbitrary units), showing the symmetrical 
spherical distribution of the old stellar tracers, obtained by summing up the Gaussian of the 
entire sample of candidate NGC~6822 stars. \label{fig:center}}
\end{figure*}
%_______________________________________________________________________________

The estimates of the center of NGC~6822 currently available date back to the seventies 
and were performed by \citet*[$\alpha_{1950} = 19^{h} 42^{m} 06^{s}.4$, 
$\delta_{1950} = -14^{\degr} 55^{'} 23^{''}$]{Gallo75}, on the basis of optical photographic 
photometry, and from \citet[$\alpha_{1950} = 19^{h} 42^{m} 04^{s}.2$, 
$\delta_{1950} = -14^{\degr} 56^{'} 24^{''}$]{Gottes77}, on the basis of radio measurements of 
neutral hydrogen across the core of NGC~6822. In Section~\ref{sec:distribution} we plan to investigate in 
detail the radial distribution of young, intermediate and old stellar tracers. 
In order to provide accurate estimates of the different radial distributions we 
performed a new and independent estimate of the center of the galaxy. To overcome possible 
systematics we decided to follow two different approaches. 

{\em i) 3D histogram}-- To overcome possible differences among different stellar tracers, we removed from the global catalog of candidate galaxy stars the young (main sequence, red supergiants) 
and the intermediate-age (AGB: C-rich, O-rich; Red Clump stars) stellar tracers. 
We only selected RGB stars and candidate red HB stars. The motivations for these 
selection criteria will become more clear in Section~\ref{sec:distribution}.  
The left panel of Fig.~\ref{fig:center}, shows the radial distribution on 
sky (RA, DEC) of older stellar tracers, together with the isocontour levels. 
Note that the radial distribution was smoothed 
by assuming an unitary Gaussian kernel with sigma equal to the sum in quadrature of the 
errors on the position of the centroid. Finally, the Gaussians of individual 
candidate galactic stars were summed up and plotted on a 3D (RA, DEC, N [arbitrary units]) 
histogram, showed in the right panel of Fig.~\ref{fig:center}. 
%________________________________________________________________________________
\begin{figure}[ht!]
\centering
\includegraphics[width=7.6cm]{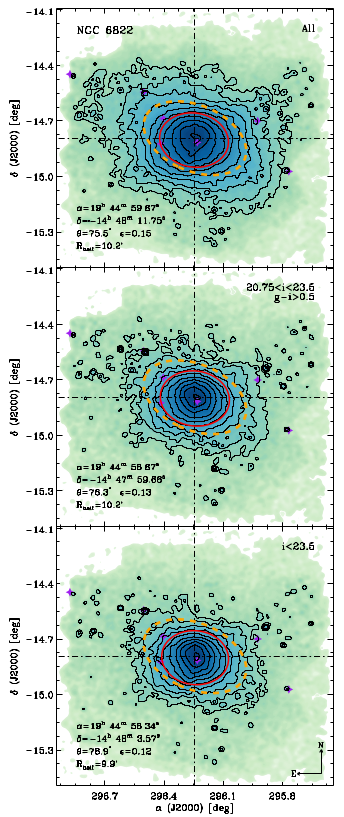}
\caption{Top: spatial distribution of the entire sample of candidate NGC~6822 stars. 
The contours display the isodensity levels from 5\% to 95\% in a logarithmic scale.
The red and orange dotted ellipses outline 
our fit and results from \citet{Zhang21}. The purple 
stars show the location of the eight known NGC~6822 GCs 
\citep[e.g.][]{Hwang11, Huxor13}.
Middle: Same as the top, but the galaxy candidates were selected both in 
magnitude (20.75 $< i <$ 23.5 mag) and in color ($g-i >$ 0.5 mag).
Bottom: Same as the middle, but the sample was only selected 
in magnitude ($i <$ 23.5 mag).
\label{fig:GlobalRadial}}
\end{figure}
%________________________________________________________________________________
%
Data plotted in this figure show that older stellar tracers attain a symmetrical spherical 
distribution over the entire body of the galaxy. We performed a Gaussian fit 
to the observed distribution and the peak of the distribution -- galaxy center -- is 
located at: R.A.(J2000) = $19^{h} 44^{m} 58^{s}.56$, DEC(J2000) = $-14^{d} 47^{m} 34^{s}.8$. The new 
estimate is in fair agreement with the optical and radio values, indeed it differs 
of 0.6 and 0.51 arcmin in RA and DEC from the estimate of \citet{Gallo75}, 
and of 1.15 and 1.53 arcmin in RA and DEC from the estimate of \citet{Gottes77}. 

{\em ii) Maximum Likelihood}-- To provide an independent estimate of the structural 
parameters of NGC~6822 we adopted the publicly available code provided by 
\citet{Martin08}. They developed a maximum likelihood algorithm to derive the 
structural parameters for a number of MW satellites. They validated the algorithm 
on SDSS data and they found that their approach provides solid estimates even for 
extremely faint stellar systems. The isocontours plotted in the top panel of 
Fig.~\ref{fig:GlobalRadial} display quite 
clearly that the spatial distribution of the galaxy is quite spherical and symmetrical. 
The coordinates of the galaxy center (see labeled values in the top panel of 
Fig.~\ref{fig:GlobalRadial} and values listed in Table~\ref{tab:param}) are in fair 
agreement with the estimate based on the 3D histogram.  
The isocontours plotted in this panel display quite clearly the presence of candidate galaxy stars
at radial distances larger than 1 degree. A similar evidence was also found by \citet{Zhang21},
but the current evidence is based on a larger sample. 
Moreover, the maximum likelihood solution suggests a position angle of $\theta \sim$ 75 degrees, 
a very modest eccentricity ($\epsilon \sim$ 0.15) and a R$_{half}$=10.2$'$.  
The current structural parameters are quite consistent with those provided 
by \citet[orange dotted ellipse in figure]{Zhang21} for the 
intermediate and old stellar tracers (RGB plus RC stars). 
They found a similar mean position angle ($\theta$ = 62 degrees), but their mean  
eccentricity is a factor of three larger (0.35 vs 0.15) and their half-light radius
is greater than our solution (16.13$'$ vs 10.20$'$).
They also provided the variation of both the position angle and the ellipticity as a function of the major-axis radius (see their Fig. 8). They found that $\theta$ decreases from 80 to 35 degrees when moving from the innermost (r = 0.2 degrees) to the outermost (r = 0.9 degrees) galactic regions, while the eccentricity in the same regions increases from 0.2 to 0.4. These findings agree quite well with results obtained by \citet*{Battinelli06} suggesting a variation in $\theta$ from 80 (r $\sim$ 0.1 degrees) to 65 degrees (r $\sim$ 0.6 degrees) and in $\epsilon$ from $\sim$ 0.20 to $\sim$ 0.38.
In passing we note that the difference in eccentricity and in half-light radius, 
based on the maximum likelihood solution, and similar estimates in the literature 
is mainly due to the depth and homogeneity of the current photometric catalog 
across the main body of the galaxy.
To constrain on a more quantitative basis the difference, we performed a new 
estimate of the structural parameters by selecting two different star samples.
The first sample was made considering all the stars with 20.75 $< i <$ 23.5 mag and $g-i >$ 0.5 mag;
the second sample was obtained by artificially cutting the photometric 
catalog to stars brighter than $i =$ 23.5 mag. The results are plotted in the central and 
bottom panels of Fig.~\ref{fig:GlobalRadial} and show that the isocontours and the
structural parameters remain very similar to the global solution. 
%Thus suggesting that the increase in the eccentricity 
%is mainly caused by young and not-too-old stellar populations. Indeed, the faint 
%limit of our photometric catalog is mainly dominated by older stellar tracers.       

%%%%%%%%%%%%%%%%%%%%%%%%%%%%%%%%%%%%%%%%%%%%%%%%%%%%%%%%%%%%%%%%%%%%%%%%%%%%%%%%%%%%%
\section{Stellar populations in NGC~6822} \label{sec:pop}
%%%%%%%%%%%%%%%%%%%%%%%%%%%%%%%%%%%%%%%%%%%%%%%%%%%%%%%%%%%%%%%%%%%%%%%%%%%%%%%%%%%%%

The CMD of NGC~6822 has been already introduced in Sections \ref{sec:data} and 
\ref{sec:separation}. We now discuss the different stellar populations present in 
the galaxy and their evolutionary and structural properties. Note that in the 
following we focus our attention only on the catalog of candidate galaxy stars 
discussed in Section \ref{sec:separation}.

%________________________________________________________________________________
\subsection{Young, intermediate and old stellar tracers} \label{sec:tracers}
%________________________________________________________________________________

\begin{figure*}[ht!]
\centering
\includegraphics[width=14cm]{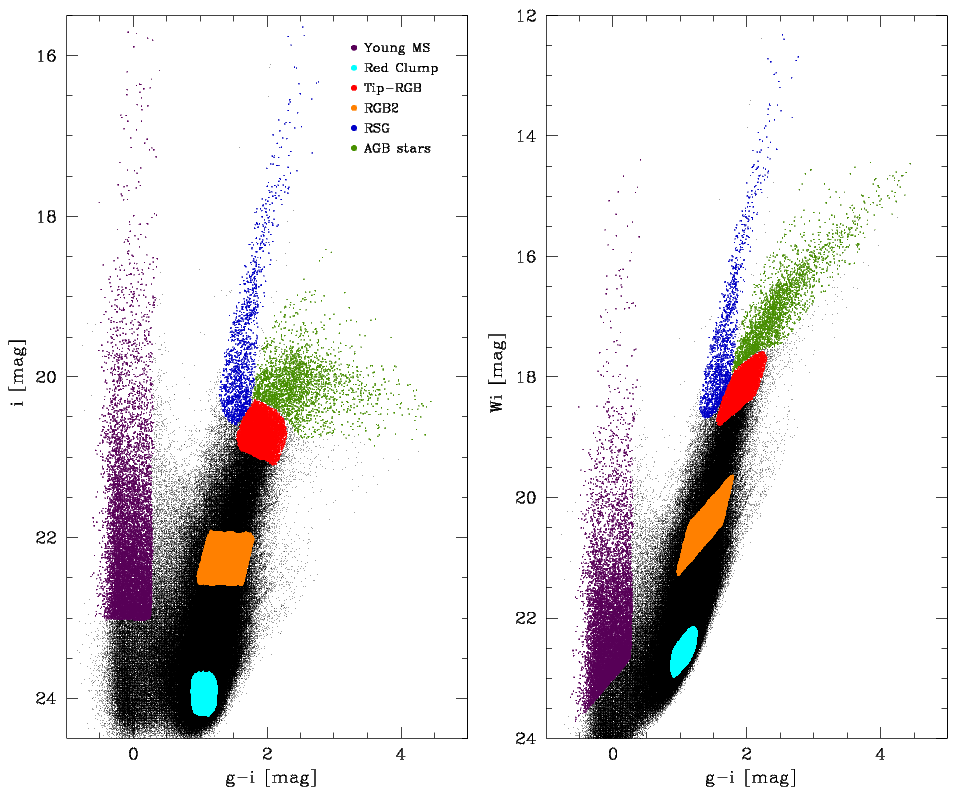}
\caption{Left: \textit{i}, \textit{g-i} CMD of candidate NGC~6822 stars. Young (MS, RSG), 
intermediate (Red Clump, AGB) and old (TRGB, RGB2) stellar tracers are marked with dots 
of different colors (see labels). 
Right: Same as the left panel, but using the 
reddening-free pseudo-magnitude called Wesenheit index 
\textit{$W_{i}$}, defined as $W_{i} = i-1.32(g-i)$ mag.  
\label{fig:optical}}
\end{figure*}

The left panel of Fig.~\ref{fig:optical} shows the \textit{i}, \textit{g-i} CMD 
of candidate NGC~6822 stars. Note that this CMD covers more than nine 
$i$-band magnitudes and shows several well-defined evolutionary phases, 
associated to young, intermediate and old stellar populations. On the basis of 
different sets of stellar isochrones covering a broad age range retrieved from 
the BASTI data base\footnote{The interested reader is referred to 
\url{http://basti-iac.oa-abruzzo.inaf.it/index.html}} 
and by visual inspection, we selected different stellar tracers, highlighted with 
dots of different colors (see labels). We focus our attention on the following 
stellar tracers. 

{\em Young stellar tracers}-- Young MS (violet dots) was selected considering stars 
with $i<$ 23 mag and -0.5 $< g-i <$ 0.3 mag. Note that the fainter limit in 
magnitude was fixed at $i = 23$ mag to avoid possible contamination of 
old and intermediate-age helium  
burning stars (HB, RC). Moreover, we also selected RSGs (blue dots) located 
between 16 $< i <$ 20.5 mag and 1.4 $< g-i <$ 2.6 mag. We selected these evolutionary 
phases because NGC~6822 is together with IC10 the prototype of star forming galaxies 
identified at high redshifts. Indeed, photometric and spectroscopic investigations 
indicate that NGC~6822 has recently experienced several star formation episodes.

%_____________________________________________________________________________________________
%_____________________________________________________________________________________________

{\em Intermediate stellar tracers}-- Red Clump (RC; cyan dots) stars were selected in 
the region located at  23.7 $< i <$ 24.3 mag and 0.8 $< g-i <$ 1.2 mag. RC stars are 
central helium burning stars and their stellar masses are intermediate between 
the low-mass (M/M$_{\odot} \le$ 1.2) stellar structures characterized by helium cores 
strongly affected by electron degeneracy and the intermediate-mass (M/M$_{\odot} \ge$ 2.3) 
stars that ignite quiescently central helium burning. The limits in stellar mass of the 
evolutionary channel producing RC stars depends on the chemical composition and their 
key feature is to ignite helium in a partially electron degenerate helium core. The 
identification of the peak associated to RC stars is based on several test and trial 
on 3D histograms (\textit{i}, \textit{g-i}, N [number of stars], \textit{r}, 
\textit{g-i}, N). Data plotted in this CMD show that RC stars overlap with RG stars. 
This degeneracy is caused by the so-called age-metallicity degeneracy along
the RGB. Indeed, RG stars associated to stellar populations covering a broad range in 
age and/or in chemical composition cover, at fixed magnitude, a broad range in colors 
over the entire RGB. Note that in the magnitude range typical of RC stars the thickness 
in \textit{g-i} color of the RGB is of the order of one magnitude. However, the 
identification of RC stars is facilitated by the empirical and theoretical evidence 
that the central helium burning lifetime of RC stars is systematically longer than 
the hydrogen-shell burning lifetime along the RGB. This means that the quoted 3D histograms 
display a well defined peak in magnitude and in color across the RC region (see Appendix~\ref{appendix1}).  

In this group we also included AGB (dark green dots) stars and they were selected as the 
stars brighter than the Tip of the RGB (TRGB, \textit{i} $\sim$ 20.3 mag) and redder 
than the RSGs. Note that AGB stars are a mixed bag, given that all stellar populations 
older than $\approx$0.5 Gyr produce AGB stars. However, in stellar systems, like NGC~6822, 
that experienced recent star formation events AGB stars are typically dominated by 
intermediate-mass (2 $\le$ M/M$_{\odot} \le$ 8) stellar structure. In passing it is worth 
mentioning that the current criteria are only able to select bright AGB stars, indeed, 
the AGB stars fainter than the TRGB (early AGB) in optical CMD overlap with RGB stars 
and they can be hardly identified.  

{\em Old stellar tracers}-- Theory and observations indicate that TRGB stars 
attain a constant \textit{i}-band magnitude over a broad range of stellar ages and 
chemical compositions. The TRGB was identified, following \citet{Sanna08} and \citet{Bono08}, by 
using both the \textit{i}-band  luminosity function and the 3D histogram 
(\textit{i}, \textit{g-i}, N) and we found that its apparent magnitude and color are 
\textit{i} $\sim$ 20.32 $\pm$ 0.01 mag, \textit{g-i} $\sim$ = 1.68 $\pm$ 0.01 mag. TRGB stars 
(red dots) were selected as the stars located between the TRGB and 0.7 mag fainter. 
We are aware that this region of the CMD includes RG stars associated to intermediate-age 
stellar populations. However their contribution, if we assume a homogeneous star formation 
rate over the age range producing RG stars and an initial mass function {\em a la Salpeter} 
(power law), is modest when compared with truly old (t $\ge$ 10 Gyr) stellar populations.
Moreover, stellar populations with a chemical composition typical of NGC~6822 ([Fe/H]= -1.05, 
\citealt{Kirby13}) stellar structures with masses larger than 2.1 M$_{\odot}$ are beyond 
the so-called red giant transition \citep*{Sweigart89, Sweigart90}. This means that 
they ignite quiescently central helium burning and the extension in magnitude and in color 
of their RG branches is quite modest when compared with truly old stellar structures.  
However, to take account of the change in the stellar populations that are distributed along 
the RGB, we also selected a fainter RGB region (RGB2, orange dots) located between the TRGB 
and the RC region: 22 $< i <$ 22.5 mag and 1 $< g-i <$ 1.6 mag. Note that the range in color 
covered by RGs in the RGB2 sample is wider than for stars in the TRGB sample, thus further 
supporting degeneracy between metallicity and age.

The criteria adopted to select the different stellar tracers rely on apparent 
magnitudes and on colors. Therefore, they are affected by possible systematics, because 
NGC~6822 is affected by differential reddening. To validate the criteria adopted for 
selecting the different populations we plotted the same stars by using the 
so-called Wesenheit index \citep{van68, Madore76, Bono10L, Bono19}. 
The reason is twofold. i) The Wesenheit index is a pseudo-magnitude that is 
reddening free by construction. This means that it is independent of reddening 
and of differential reddening. ii) It can be derived by using measurements in 
two different bands.  However, the Wesenheit magnitude is also affected by two 
main drawbacks: i) it relies on the assumption that the reddening law is universal; 
ii) the intrinsic error is larger than typical error affecting single band magnitudes. 
We defined the i-band Wesenheit magnitude by using the reddening law provided by 
\citet{Cardelli89} and the central wavelengths of the Pan-STARRS photometric 
system \citep{Tonry12}. In particular, we found $W_{i} = i-1.32(g-i)$ mag. 
The right panel of Fig.~\ref{fig:optical} shows the stellar populations 
selected in the left panel and with the same color coding. Note that in this CMD the 
reddening and the differential reddening can only affect the spread in color. 
Data plotted in this panel show that the use of the wesenheit index, taking account 
for the color information, further improve the definition of several evolutionary 
phases. Candidate AGB stars cover more than three Wesenheit magnitudes and they 
are distributed along a well defined and narrow sequence. The RSG sequence is 
narrower than in the \textit{i}, \textit{g-i} CMD and fully supports the cuts 
in magnitude and colors adopted to separate RSG and AGB stars.   
The boxes adopted for selecting TRGB, RGB2 and RC stars are, as expected, 
slanted towards brighter magnitudes. Redder stars, at fixed magnitude, 
become systematically brighter in Wesenheit magnitude. As a whole they display 
similar features, thus suggesting that differential reddening plays a minor 
role in the selection of the different stellar tracers.

%________________________________________________________________________________
\subsubsection{Near infrared and mid infrared magnitudes} \label{sec:infrared}
%________________________________________________________________________________

The Wesenheit magnitude allowed us to constrain the impact that reddening and 
differential reddening have on the magnitude adopted for selecting the 
different stellar tracers. However, the \textit{g-i} color is almost a factor 
of two more sensitive to reddening uncertainties than typical optical colors  
(E(\textit{B-V})/E(\textit{g-i})$\sim$0.63; \citealt{Cardelli89}). 
%________________________________________________________________________________
\begin{figure}[ht!]
%\centering
\includegraphics[width=9.1cm]{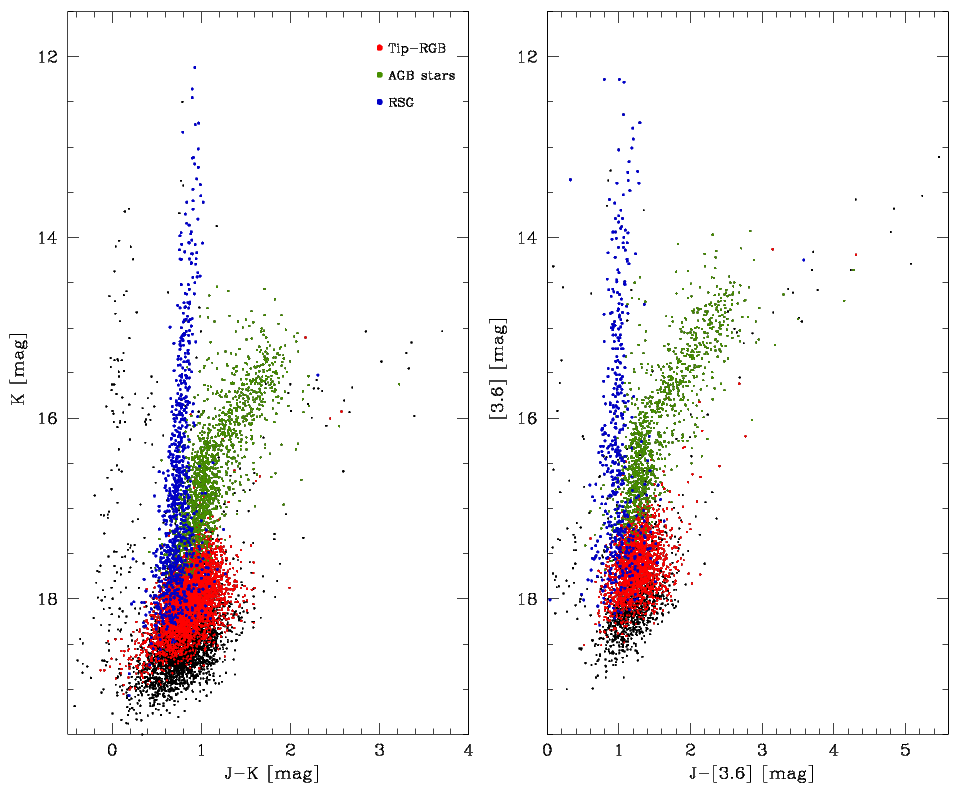}
\caption{Left: NIR \textit{K}, \textit{J-K} CMD for the bright stellar populations
(RSGs, AGB, TRGB)  in NGC~6822. The color coding is the same as in 
Fig.~\ref{fig:optical}.  
Right: Same as the left, but for the NIR-MIR \textit{[3.6]}, \textit{J-[3.6]} CMD. 
\label{fig:IR}}
\end{figure}
%________________________________________________________________________________
%
To investigate on a more quantitative 
basis the reddening impact on the selection criteria we decided to take advantage 
of the NIR and MIR catalogs introduced in Section~\ref{sec:data}.
The NIR dataset includes photometric measurements in $J,H,K$ bands down to limiting 
magnitudes of J$\sim$20.6 mag, H$\sim$20.0 mag and K$\sim$19.6 mag (\citetalias{Sibbons12}).  
The MIR photometry based on SPITZER images \citep{Khan15} includes measurements in 
the IRAC-bands and their limiting magnitudes are either similar 
or slightly brighter than the NIR catalog. We cross correlated the 
three different catalogs and we ended up with an optical-NIR-MIR catalog including 
$\sim$97,500 stars with at least one optical and one NIR magnitude and with 
$\sim$30,023 stars with at least one optical and one MIR magnitude.
%________________________________________________________________________________
\begin{figure}[ht!]
%\centering
\includegraphics[width=9.1cm]{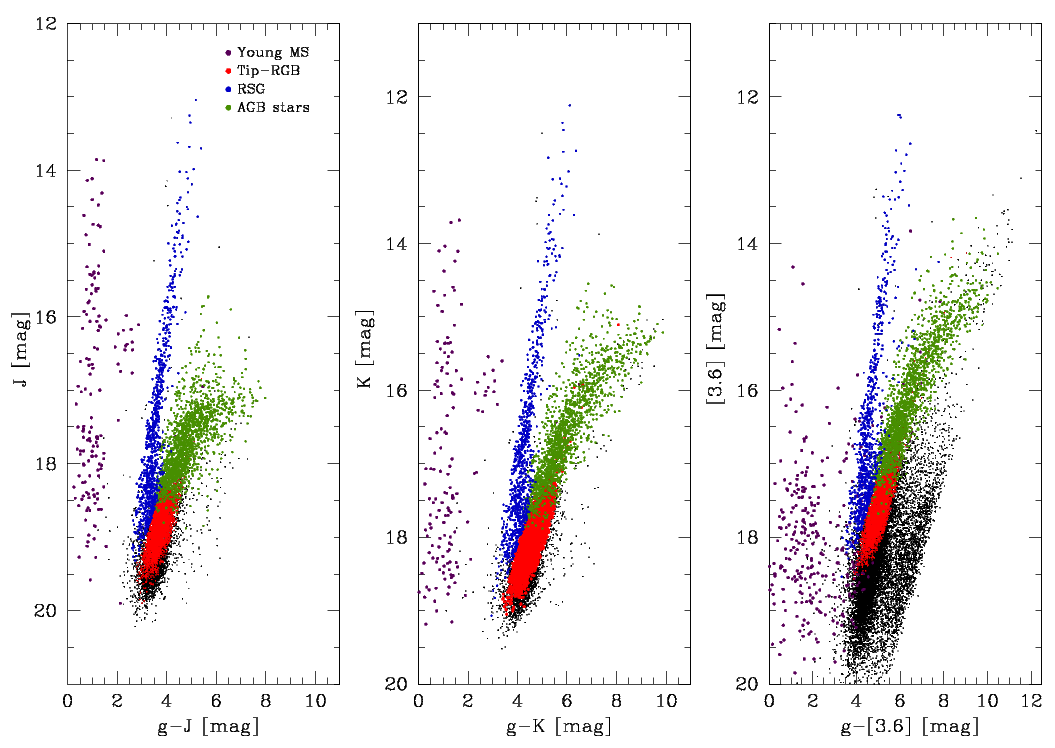}
\caption{Same as Fig. \ref{fig:optical}, but young (MS, RSG), intermediate 
(AGB) and old (TRGB) stellar tracers are plotted in the two optical-NIR 
\textit{J}, \textit{g-J} (middle), and \textit{K}, \textit{g-K} (middle) CMDs  
and in the optical-MIR \textit{[3.6]}, \textit{g-[3.6]} CMD (right).
The color coding is the same as in Fig. \ref{fig:optical}. 
\label{fig:validation}}
\end{figure}
%________________________________________________________________________________

Data plotted in the left panel of Fig.~\ref{fig:IR} show the NIR \textit{K}, \textit{J-K} CMD  
of the bright stellar populations in NGC~6822. The evolutionary sequences for RSGs 
and AGB are quite narrow in color and cover a wide range in magnitudes. Moreover, 
RSGs and AGB stars display a clear separation for K $\le$ 16.5 mag. However, TRGB stars
are distributed over a range of two magnitudes in K and one magnitude in J-K and 
overlap with RSGs and AGB stars. This means that the separation among the three different 
stellar tracers in this NIR CMD is prone to possible systematics. The right panel of the 
same figure shows the NIR-MIR \textit{[3.6]}, \textit{J-[3.6]} CMD for the same stellar 
tracers. The separation between RSGs and AGB stars becomes more clear down to 
\textit{[3.6]} $\sim$ 17 mag, but the mix of the three different stellar tracers at fainter 
magnitudes is still present.    

Data plotted in NIR and in NIR-MIR CMDs are hampered by the limited sensitivity of these 
colors to the effective temperature. The \textit{g-i} color is more sensitive to the 
effective temperature than NIR and NIR-MIR colors, but it is significantly smaller than 
optical-NIR and optical-MIR colors. Indeed, the difference in color between RSG, TRGB 
and bright AGB stars is of the order of three magnitudes in \textit{g-i}, but it becomes 
of the order of seven magnitudes in \textit{g-K} and of eight magnitudes in \textit{g-[3.6]}.     
From left to right, Fig.~\ref{fig:validation} shows the \textit{J}, \textit{g-J} 
(left) and the \textit{K}, \textit{g-K} (middle) optical-NIR CMDs, while the right panel 
shows the  \textit{[3.6]}, \textit{g-[3.6]} optical-MIR CMD. Note that the CMDs 
plotted in this figure only cover bright stellar populations, indeed RC, RGB2 
and a significant fraction of young MS (violet dots) stars are too faint in NIR 
and in MIR bands. However, brighter and cooler stellar populations (TRGB, RSG, AGB) 
display narrow, well defined sequences and cover a very broad range in magnitudes 
and in colors. 

Data plotted in this figure display several interesting 
features worth being discussed in more detail. 
i) AGB and TRGB stars overlap over a substantial range in magnitude ranging 
from 0.3-0.4 mag in the \textit{J}-band, to 0.5-0.6 mag in \textit{K} and to 
0.8-0.9 mag in \textit{[3.6]}. This means that photometric selections only 
based on optical/NIR/MIR CMDs are, once again, prone to systematics, since 
TRGB and AGB stars cover similar magnitude and colors. The same outcome 
applies to the separation between RSGs and TRGB stars.
ii) AGB stars cover more than three magnitudes in \textit{[3.6]}, and six 
magnitudes in \textit{g-[3.6]} color, thus providing a firm validation for the 
separation of cool and warm AGB stars.  
iii) Young MS stars (violet dots) are also present, but only the very bright massive 
stars have been detected.  
The current empirical evidence indicates that the optical/NIR/MIR CMDs concerning 
young and intermediate-age stellar tracers bring forward several advantages (reduced 
sensitivity to reddening and differential reddening, strong sensitivity to effective 
temperature) when compared with NIR and optical CMDs, but they are affected 
by systematics when dealing with old stellar tracers.

%________________________________________________________________________________
\subsection{AGB: selection of C-rich and O-rich stars} \label{sec:AGB}
%________________________________________________________________________________

%________________________________________________________________________________
\begin{figure*}
\centering
\includegraphics[width=1.4\columnwidth]{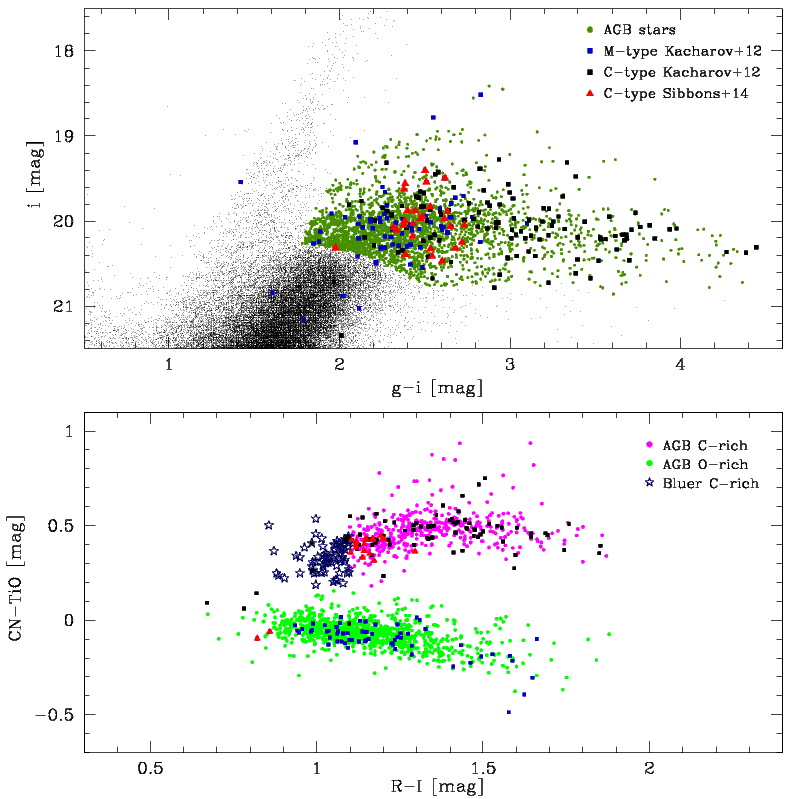}
%\plotone{AGB_sample.png}
\caption{Top: \textit{i}, \textit{g-i} CMD of candidate AGB stars in NGC~6822. 
The CMD is zoomed on the bright portion to show the current AGB sample 
(dark green dots). Blue squares and red triangles 
display spectroscopically confirmed C-type (\citetalias{Kacharov12}, \citetalias{Sibbons15}), 
while black squares show confirmed M-type stars (\citetalias{Kacharov12}). 
Bottom: \textit{CN-TiO}, \textit{R-I} color-color diagram for candidate AGB stars plotted 
in the top panel. The diagram is based on multi-band photometry collected by 
\citetalias{Letarte02} and shows a clear separation between candidate C-rich (magenta dots) 
and O-rich (green dots) stars. The blue star symbols highlight the so-called bluer C-stars,
as defined by \citetalias{Letarte02}. Symbols and colors of the spectroscopic sample (C-, M-type) 
are the same as the top panel.
\label{fig:AGBsample}}
\end{figure*}
%________________________________________________________________________________
%
We focused our attention on AGB stars because they cover a broad range in stellar 
masses (M$\sim$1--8~M$_{\odot}$), and in turn, stellar ages (from a few hundred Myrs 
to more than 10 Gyrs). This means that their spatial distributions can be safely 
adopted to trace back in time and in space the star formation episodes experienced 
by stellar systems. 
The AGB stars are classified according to their carbon-to-oxygen  
abundance ratios in three different groups: C-type (C/O$>$1), O-type (C/O$<$1, 
called also M-type) and S-type (C/O$\sim$1). More recently have been suggested 
several subclasses and the reader interested in a more detailed discussion is 
referred to \citet{Boyer11, Boyer15}. 
Photometric and spectroscopic investigation 
indicate that C- and O-type stars typically display different radial 
distributions with the latter ones more centrally concentrated than the former 
ones. This means that a comprehensive analysis of O-rich and C-rich AGB stars 
covering the entire body of the galaxy will provide solid constraints on their 
dependence on the environment and on their metallicity distribution.    

We have already mentioned in Section~\ref{sec:tracers} that candidate AGB stars were selected in the 
\textit{i}, \textit{g-i} CMD (see Fig.~\ref{fig:optical}) as the stars brighter 
than the TRGB (\textit{i} $\sim$ 20.3 mag) and redder than the RSGs. The top panel 
of Fig.~\ref{fig:AGBsample} shows in the same CMD a zoom in of the entire sample 
of candidate AGB stars (2788 stars, dark green dots). To further support the 
photometric selection of AGB stars we took advantage of the spectroscopic catalogs 
of C- and M-type stars available in the literature for NGC~6822. A detailed 
spectroscopic investigation was performed by \citetalias{Kacharov12}. 
They used low-resolution spectra collected with the 
multi-object slit spectrograph VIMOS at VLT (ESO) and identified 150 C-type 
stars and 122 M-type stars. A similar spectroscopic analysis was also performed 
by \citetalias{Sibbons15} who provided a sample of 82 C-type and one anomalous M-type 
giant by using the AAOmega multi-fibre spectrograph at the Anglo-Australian 
Telescope (AAT). 
The spectroscopic catalogs were cross-matched with our optical catalog and the 
objects in common are plotted in the top panel of Fig.~\ref{fig:AGBsample}. 
The C-type identified by \citetalias{Kacharov12} are marked with black 
squares, while those identified by \citetalias{Sibbons15} as red triangles; the 
M-type identified by \citetalias{Kacharov12} are plotted as blue squares. 
A glance at the data plotted in this panel display that candidate AGB stars 
selected by using photometric properties agree quite well with spectroscopically 
selected samples. The photometric sample, as expected, outnumbers the spectroscopic 
sample, but they cover similar  magnitude and color ranges. Note that spectroscopic 
M-type stars attain, as expected, colors that are systematically bluer than C-type 
stars. However, the two subgroups are degenerate in the \textit{i}, \textit{g-i} 
CMD because they display a clear separation neither in magnitude nor in color.     
In passing we also note that we investigated the 
few outliers located in regions typical either of the RGs or of the RSGs. We 
found that they have close companions and the cross-match was not univocal i.e. 
there is the presence of multiple stars within the typical searching radius (1~arcsec).  
 
%____________________________________________________________________________
\begin{figure*}
\plotone{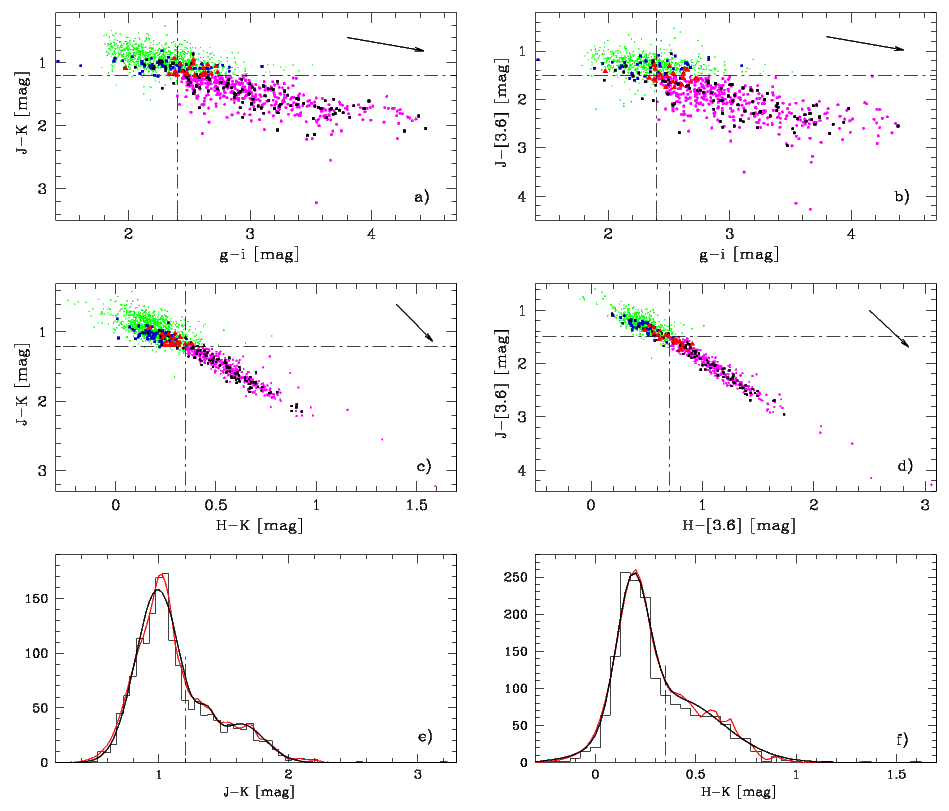}
\caption{Panel a): Optical-NIR (\textit{J-K}--\textit{g-i}) CCD for candidate AGB stars plotted in the top panel of Fig.~\ref{fig:AGBsample}. The dot-dashed lines show the color cuts adopted to separate C-rich (magenta dots) and O-rich (green dots). We performed a fit of the different color distributions following the same approach adopted by \citet{Kang06}. See text for more details. Symbols and colors of the spectroscopic sample (C- and M-type) are the same as in Fig.~\ref{fig:AGBsample}. The black arrow plotted in the top right corner displays the reddening vector for an arbitrary extinction value. 
Panel b): Same as panel a), but for optical-NIR-MIR (\textit{J-[3.6]}--\textit{g-i}) CCD. Note that the color cuts are different. 
Panel c): Same as panel a), but for NIR (\textit{J-K}--\textit{H-K}) CCD. 
Panel d): Same as panel a), but for NIR-MIR (\textit{J-[3.6]}--\textit{H-[3.6]}) CCD.
Panel e): Histogram in the NIR color J-K of candidate O- and C-rich stars. The thin red line shows the same color distribution, but smoothed with a Gaussian kernel with unitary weight and $\sigma$ equal to the photometric errors (summed in quadrature) on individual color measurements. The black line shows the multi-Gaussian (three) fit of the smoothed color distribution. The dot-dashed line shows the color cut at \textit{J-K} = 1.2 mag. 
Panel f): Same as the panel e), but for \textit{H-K} color distribution. Note that the Gaussian fit was performed by using two Gaussians. The color limit is at \textit{H-K} = 0.35 mag. 
\label{fig:AGBcanonici}}
\end{figure*}
%____________________________________________________________________________

%____________________________________________________________________________
\begin{figure*}
\plotone{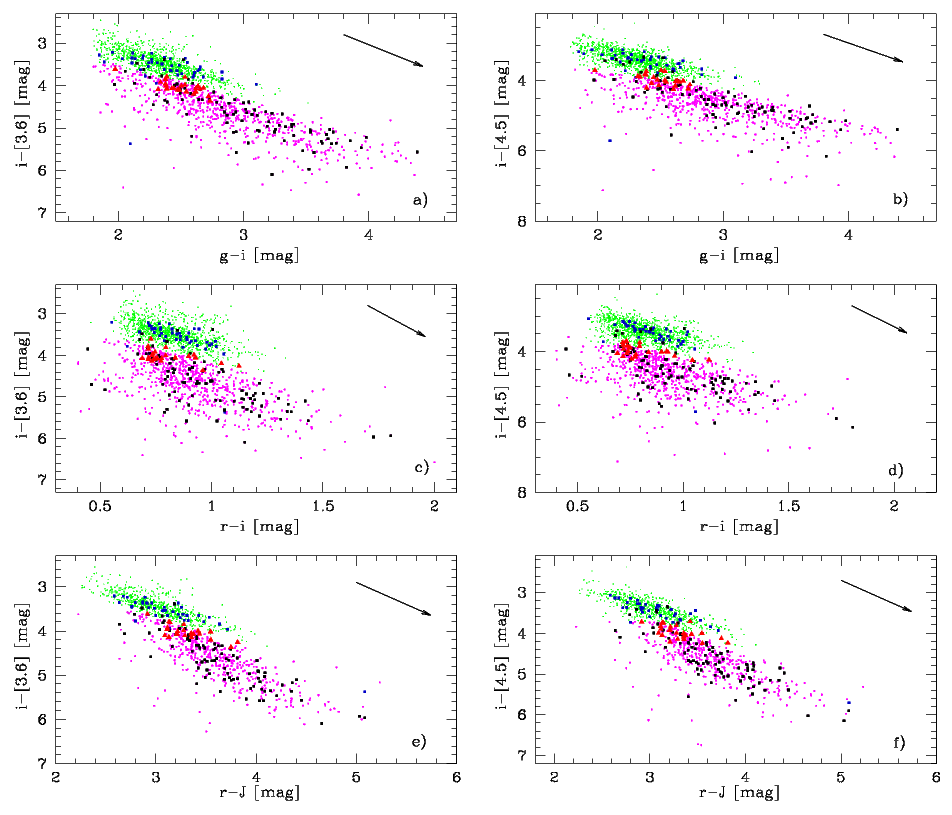}
\caption{Panel a): Same as panel a) in Fig.~\ref{fig:AGBcanonici}, but for optical-MIR (\textit{i-[3.6]}--\textit{g-i}) CCD. Note that C-rich and O-rich candidates are distributed along two different, almost parallel, sequences. The C-rich candidates cover more than three magnitudes in the optical-MIR, \textit{i-[3.6]}, color and $\sim$ 2.5 mag in the optical, g-i, color. The range in colors covered by O-rich candidates is roughly a factor of two smaller. Symbols and colors of the spectroscopic AGB sample are the same as in Fig.~\ref{fig:AGBsample}. The black arrow plotted in the top right corner displays the reddening vector for an arbitrary extinction value. 
Panel b): Same as panel a), but for optical-MIR (\textit{i-[4.5]}--\textit{g-i}) CCD. 
Panel c): Same as panel a), but for optical-MIR (\textit{i-[3.6]}--\textit{r-i}) CCD. 
Panel d): Same as panel a), but for optical-MIR (\textit{i-[4.5]}--\textit{r-i}) CCD. 
Panel e): Same as panel a), but for optical-NIR-MIR (\textit{i-[3.6]}--\textit{r-J}) CCD. 
The key advantage of this CCD (the same outcome applies to the CCD plotted in Panel f)) 
when compared with optical-MIR CCDs is that C-rich and O-rich display different slopes. 
Panel f): Same as panel a), but for optical-NIR-MIR (\textit{i-[4.5]}--\textit{r-J}) CCD. 
\label{fig:AGBcolor}}
\end{figure*}
%____________________________________________________________________________

Spectroscopic samples of C- and O-rich stars are based on solid diagnostics, but 
they are limited to the central regions of a stellar system and quite often 
hampered by modest statistics. To overcome these limitations we decided to 
use the sample of candidate C- and O-rich stars provided by \citetalias{Letarte02}.  
The key advantage of their approach is that they use a mix of broad (R, I) and 
narrow (CN, TiO) band photometry for the identification of candidate C- and O-rich 
stars. In their investigation on AGB stars in NGC~6822 they used multi-band 
photometry collected with both the Swope telescope and the wide field imager 
CFH12K at CFHT. They identified 904 candidate C-type stars, the sample was selected 
by using sharp cuts in the color-color plane: (R-I)$_{0}$ $>$ 0.90 mag and 
(CN-TiO) $>$ 0.3 mag. Note that to unredden the R-I color they used a mean 
reddening of $E(R-I)$=0.20 mag. 
Although this CCD is a very solid diagnostic for the identification of 
C-type stars, the identification of O-rich stars is hampered by the contamination 
of field red giants, because they attain similar colors (see Fig.~3 in \citetalias{Letarte02}).
We have already discussed in Section~\ref{sec:separation} the approach we adopted to separate candidate 
field and galaxy stars and to fully exploit its potential we cross-correlated their catalog, 
kindly provided by the authors in electronic form, with our sample of candidate AGB stars.
The results of the cross match are shown in the bottom panel of Fig.~\ref{fig:AGBsample}. 
Once contaminant field stars are removed, the \textit{CN-TiO}, \textit{R-I} CCD 
becomes a very solid diagnostic to identify both C-rich (486, magenta dots) and O-rich (1002, green dots). 
The blue star symbols in figure highlight the so-called 'Bluer C stars', as defined by 
\citetalias{Letarte02} (0.8 $< (R-I) <$ 1.1), that we excluded from our sample of 
candidate C-rich because they deserve a more detailed spectroscopic analysis.
Note that the number of candidate C-type stars is significantly smaller than the original sample 
provided by \citetalias{Letarte02} due to difficulties either in the astrometric solution of 
the two different data sets or in separating candidate field and galaxy stars. 
To further validate the current selection of AGB stars we have also plotted the spectroscopically 
confirmed C- and M-type stars (same as the top panel of Fig.~\ref{fig:AGBsample}). 
Spectroscopic data plotted in this CCD based on broad and narrow band photometry 
clearly show the separation between candidate C- and O-rich stars.  

To further investigate this longstanding problem of the identification of 
C-rich and O-rich stars we decided to take advantage of the multi-band catalog 
we built up. In this context it is worth mentioning that \citet{Kang06} 
have defined a solid photometric approach for selecting candidate C- and 
O-rich stars by using NIR photometry. In their investigation on NGC~6822 
they used the \textit{J-K}, \textit{H-K} CCD as a diagnostic to identify 
C- and M-type stars. They showed that the histograms in NIR colors 
(\textit{J-K}, \textit{H-K}) of candidate AGB stars have a main peak 
associated to M-stars and a long red tail associated to C-stars. Following 
this approach and the NIR color cuts (\textit{J-K}=1.53 mag, \textit{H-K}=0.5 mag) 
provided by \citet{Davidge05}, they provided a solid separation between 
C- and M-type stars in NGC~6822 (see their Fig.~7). 
To take advantage of the current optical/NIR/MIR catalog we decided to follow 
a similar approach, but with different criteria to define the color cuts. Data 
plotted in Fig.~\ref{fig:AGBcanonici} show selected AGB stars on different 
optical-NIR (\textit{J-K}--\textit{g-i}, panel a)), 
NIR (\textit{J-K}--\textit{H-K}, panel c)) and 
optical-NIR/MIR (\textit{J-[3.6]}--\textit{g-i}, 
\textit{J-[3.6]}--\textit{H-[3.6]}, panels b) and d)) CCDs. 
For each CCD we derived histograms in color (thin grey lines in 
panels e) and f)) and performed a multi-Gaussian fit (black lines) of the smoothed 
distribution (red solid lines in panels e) and f)). The color distribution 
was smoothed with a Gaussian kernel with unitary weight and $\sigma$ equal 
to the photometric errors (summed in quadrature) on individual color 
measurements. We performed a series of test and trials to define objective 
criteria for the separation of candidate C- and M-type stars, and eventually 
we decide to fix the color cuts as the mean between the peaks of primary and 
secondary Gaussian (black lines in panels e) and f), independently of 
the number of Gaussians used to fit the global distribution in color. 
We found the following color cuts: \textit{J-K}= 1.2, \textit{J-[3.6]}= 1.5, 
\textit{g-i}= 2.4, \textit{H-K}= 0.35 and \textit{H-[3.6]}= 0.7 and they are 
plotted as dashed-dotted lines in the four top panels of Fig.~\ref{fig:AGBcanonici}. 
The approach outlined in this section appears quite promising, indeed, the
color sequences and the reddening vectors (black arrows) display different 
slopes. This suggesting that the variation in color are intrinsic and only 
minimally affected by differential reddening. However, this approach is 
still affected by thorny problems. 
i) The NIR, optical/NIR and optical/NIR/MIR CCDs showed in this figure display well 
defined color sequence when moving from M- to C-type stars. This means that they 
do not show any clear separation, as in the (CN-TiO), (R-I) CCD, and/or a change in 
the slope. This evidence is further supported by the comparison with the 
spectroscopic samples. Indeed, several spectroscopic C-type stars attain colors 
that are more typical of M-type stars and viceversa. 
ii) The main peak associated to M-type stars is affected by the contamination 
of field stars, since they typically have similar colors. 
%____________________________________________________________________________
\begin{figure}
%\centering
\includegraphics[width=8.7cm]{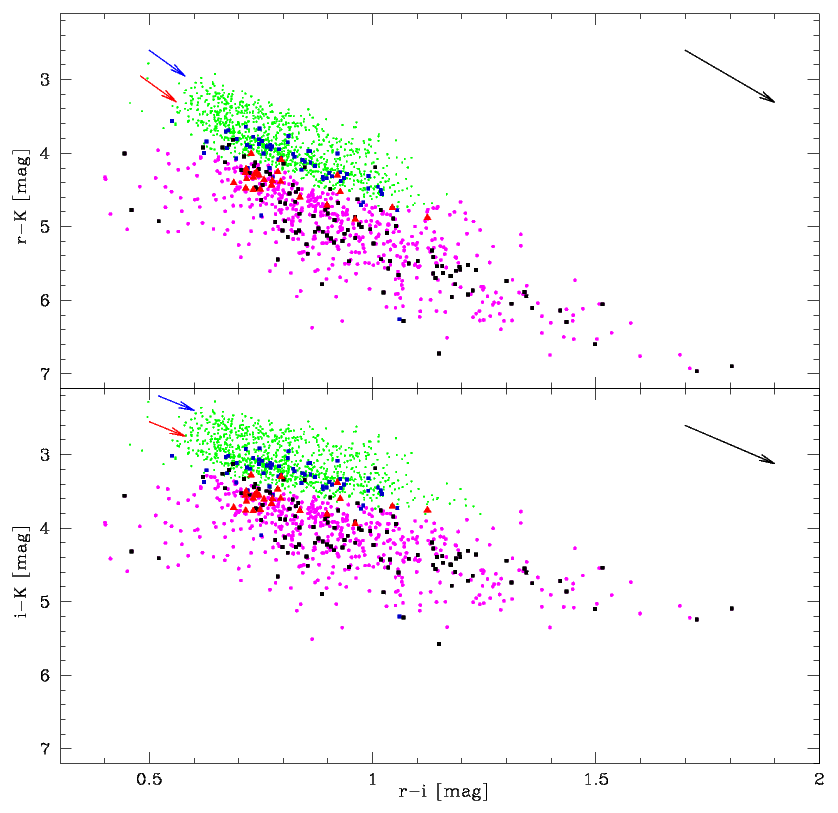}
\caption{Top: Same as panel a) in Fig.~\ref{fig:AGBcanonici}, but for optical-NIR (\textit{r-K}--\textit{r-i}) CCD. The red arrows plotted in the top left corner display the
separation in two different sequences of candidate O-rich stars. Symbols and colors of the spectroscopic AGB sample are the same as in Fig.~\ref{fig:AGBsample}. The black arrow plotted in the top right corner displays the reddening vector for an arbitrary extinction value. 
Bottom: Same as the top panel, but for optical-NIR (\textit{i-K}--\textit{r-i}). Note that in this CCD is still visible the separation of O-rich candidates into two different sequences.
\label{fig:Oseparation}}
\end{figure}
%____________________________________________________________________________
%
This means that a solid separation between field and cluster stars is required as preliminary 
step to the definition of the color cuts.

To overcome these difficulties we decided to investigate other 
possible color combinations among the available optical, NIR and MIR magnitudes 
for NGC~6822 to search for CCDs that would allow us to better identify C- and 
M-stars. A summary of this effort is showed in Fig.~\ref{fig:AGBcolor}. 
Data plotted in the optical/MIR CCDs (panels a,b,c,d) clearly show that C- and M-type 
stars are distributed along two, almost parallel sequences. The main difference 
between the top and the middle panels is that the spread in optical color is, 
at fixed optical/MIR color (\textit{i-[3.6]}, left panels; 
\textit{i-[4.5]}, right panels), significantly smaller in \textit{g-i} than in 
\textit{r-i}. Note that the separation is fully supported by the spectroscopic 
samples that are distributed along the expected sequences. Moreover, the C-rich 
sample covers more than three magnitudes in optical-MIR colors, and from 1.5 
(\textit{r-i}) to 2.5 (\textit{g-i}) mag in optical colors. The O-rich sample 
covers ranges in optical-MIR and in optical colors that are a factor of two smaller. 

The same outcome applies to the optical-NIR-MIR (\textit{i-[3.6]}--\textit{r-J} 
and \textit{i-[4.5]}--\textit{r-J}) CCDs plotted in the bottom panels of 
Fig.~\ref{fig:AGBcolor}; panels e), f)) CCDs. The key advantage of these CCDs is that sequences 
associated to O- and C-rich stars display for the first time different slopes. 
Indeed, the use of an optical/NIR color, together with an 
optical/MIR color, causes the C-rich sequence to be more slant when compared 
to optical colors (top, middle panels). Note that optical/NIR colors cover ranges 
roughly similar to the optical-MIR colors. It is worth mentioning that 
the C-rich and the O-rich sequences attain in the quoted CCDs slopes that differ 
from the slope of the reddening vectors (black arrows). This means that the 
dispersion in color present in all the selected CCDs is an intrinsic feature 
of the two stellar populations.

%_______________________________________________________________________________
\begin{figure*}[ht!]
\centering
\includegraphics[width=14cm]{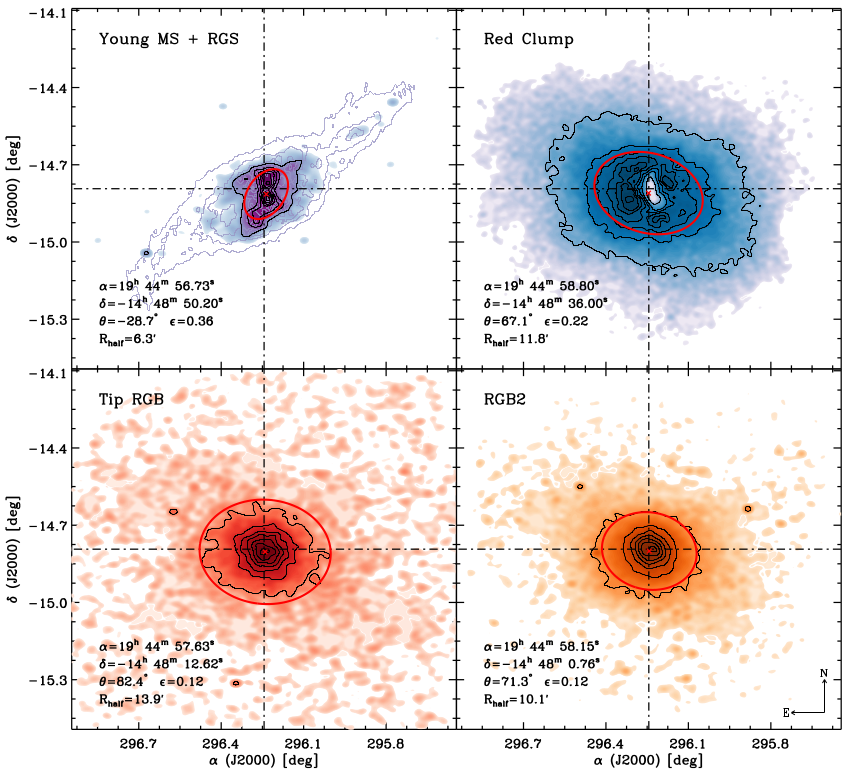}
\caption{Spatial distributions of the young (Young MS + RSG; top-left panel), intermediate (RC; top-right panel) and old (TRGB, RGB2; bottom panels) stellar tracers in NGC~6822.
The contours show the isodensity levels from 5\% to 95\% in a logarithmic scale. 
The grey contours plotted in the top-left panel (young tracers) highlight the isocontours
of the HI column density map from \citet{deBlok00, deBlok03, deBlok06}.
The dot-dashed lines trace the estimated galaxy center (see Section \ref{sec:center}). The thick red ellipses outline the fit performed according to the maximum likelihood algorithm by \citet{Martin08}. The derived structural parameters are labeled in figure and listed in Table \ref{tab:param}.  
\label{fig:radial}}
\end{figure*}
%_______________________________________________________________________________

The mix of optical/NIR/MIR colors showed in Fig.~\ref{fig:AGBcolor} brought forward several key  
advantages in the identification of C- and M-type stars. Therefore, we decided 
to investigate the use of these colors to further characterize the sample of 
C-rich and O-rich stars. Interestingly enough we found that CCDs based on two 
optical-NIR colors (\textit{r-K}--\textit{r-i}, \textit{i-K}--\textit{r-i}, 
top and bottom panel of Fig.~\ref{fig:Oseparation}) display the same features 
of optical/NIR/MIR CCDs, but they also highlight a separation in two parallel 
sequences  of candidate O-rich stars (see the blue and the red arrow plotted 
on the top left corner of the two panels). Note that spectroscopically confirmed 
O-rich stars are distributed along the two sequences, but the bluer sequence 
only includes a few spectroscopic identification. This empirical evidence indicates 
that the O-rich stars in NGC~6822 appear to include two different stellar components.
Plain physics arguments suggest that the bluer sequence is younger, while the redder 
is older. More quantitative constraints require a detailed comparison between 
theory and observations. 

The current findings indicate that either optical-NIR (Fig.~\ref{fig:Oseparation}) 
or optical-NIR-MIR (Fig.~\ref{fig:AGBcolor}) CCDs including either the r- or the 
i-band allow us a solid identification and characterization of C-rich and O-rich.

%________________________________________________________________________________
\subsection{Radial distributions of the different stellar tracers} \label{sec:distribution}
%________________________________________________________________________________

%_______________________________________________________________________________
\begin{figure*}[ht!]
\centering
\includegraphics[width=15cm]{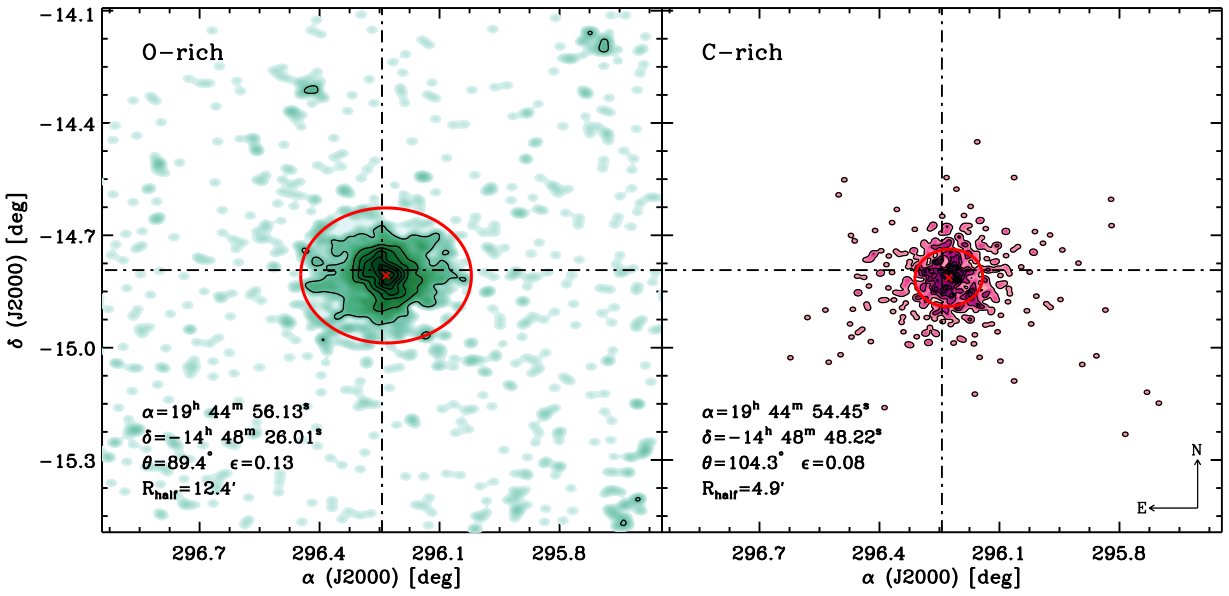}
\caption{Same as Fig.~\ref{fig:radial}, but for O- (left panel) and C-rich (right panel) stars. C-rich stars are based on the selection performed on the \textit{r-K}--\textit{r-i} CCD, while the O-rich stars come from our global selection of AGB stars (\textit{i}, \textit{g-i} CMD). 
\label{fig:OeCradial}}
\end{figure*}
%_______________________________________________________________________________

The selection of stellar tracers, covering a broad range of stellar ages, provide 
the opportunity to investigate on a quantitative basis their radial distributions. 
To provide quantitative constraints on their structural parameters we adopted 
the same algorithm we applied to the global catalog in Section~\ref{sec:center}. 
The radial distributions and the isodensity contours of the selected stellar populations 
are shown in Fig.~\ref{fig:radial} and in Fig.~\ref{fig:OeCradial}, 
the results of the fit are overplotted as thick red ellipses and the structural 
parameters are also labeled and listed in Table~\ref{tab:param}. 

The top left panel of Fig.~\ref{fig:radial} shows the radial distribution on sky 
of young stellar tracers: MS and RSGs.
A glance at the isocontours plotted in the top left panel clearly show that young 
stellar tracers are distributed along a well-defined bar that is mainly located 
between the first and the second quadrant, bending into the fourth quadrant in 
its southern/southern-east extension. Moreover and even more importantly, the 
young bar is systematically off-center (see black dashed-dotted lines and red 
cross) by 0.46 arcmin in RA and 1.26 arcmin in DEC. The spatial distribution of 
young stellar tracers is very similar to the distribution obtained by 
\citet{Hirsc20} for their RSG sample.
The solution of the fit shows a moderate eccentricity, its position angle is  
$\theta \sim$ -29 degrees, i.e. mainly distributed along the second-fourth quadrant, 
and it has a modest half-light radius (R$_{half} \sim$ 6.3$'$). The 
isocontours are also suggesting that the position angle is driven by young stars 
located outside the bar, thus suggestive of the possible presence of a disk.
This working hypothesis is also supported by the isocontours of the HI column density map
obtained by \citet{deBlok00, deBlok03, deBlok06} and overplotted on the sky distribution of young stars.   

The radial distribution of intermediate-age stellar tracers (RC) when compared with 
young stellar tracers is significantly different (see top right panel of Fig.~\ref{fig:radial}). 
The isocontours are far from being symmetric and in the innermost regions 
there is evidence of a steady decrease of RC stars along the bar made by young stars. 
This is an observational bias. The RC stars are relatively faint ($i \sim$ 24 mag) and 
the innermost galactic regions are more crowded and more affected by internal 
reddening. This means that our optical catalog is not complete in the innermost 
regions. To overcome this limitation we estimated the peak of the RC distribution 
by performing a Gaussian fit of the RC distribution in RA and in DEC (see red cross). 
Eventually, we performed the fit, at fixed center position, and the results are 
plotted in the same panel.   
The position angle is $\theta \sim$ 67 degrees, i.e. mainly distributed along the 
first-third quadrant. Moreover, the solution of the fit indicates a smaller 
eccentricity ($\epsilon \sim$ 0.2) and its half-light radius is almost a factor 
of two larger. This means that RC stars are distributed over a significant fraction 
of the body of the galaxy.     

%_______________________________________________________________________________________
\begin{deluxetable*}{lllccccc}
\tablenum{4}
\tablecaption{Structural parameters for the selected stellar tracers. From left to right 
the columns give: the stellar tracer, sky coordinates of the galaxy center, 
the position angle, the eccentricity, the half-light radius, difference in RA and DEC with 
the galaxy center estimated in Section~\ref{sec:center} (see also Fig.~\ref{fig:radial} and \ref{fig:OeCradial}). \label{tab:param}}
\tablewidth{0pt}
\tablehead{
\colhead{Stellar tracer} & \colhead{$\alpha$} & \colhead{$\delta$} & \colhead{$\theta$} & \colhead{$\epsilon$} & \colhead{R$_{half}$} & \colhead{$\Delta$RA} & \colhead{$\Delta$DEC} \\
\colhead{} & \colhead{(J2000)} & \colhead{(J2000)} & \colhead{($\degr$)} & \colhead{} & \colhead{($'$)} & \colhead{($'$)} & \colhead{($'$)}
} %\decimalcolnumbers
\startdata
Global$^{a}$ & $19^{h} 44^{m} 58.56^{s}$ &\ $-14^{h} 47^{m} 34.80^{s}$ & $\ldots$ & $\ldots$ & $\ldots$ & $\ldots$ & $\ldots$ \\
G1$^{b}$     & $19^{h} 44^{m} 58.56^{s}$ &\ $-14^{h} 47^{m} 34.80^{s}$ & 75.8     & 0.14     & 10.2     & $\ldots$ & $\ldots$ \\
G2$^{c}$     & $19^{h} 44^{m} 59.67^{s}$ &\ $-14^{h} 48^{m} 11.75^{s}$ & 75.5  & 0.15 & 10.2 & -0.28 & 0.62 \\
G3$^{d}$     & $19^{h} 44^{m} 58.67^{s}$ &\ $-14^{h} 47^{m} 59.66^{s}$ & 76.3  & 0.13 & 10.2 & -0.03 & 0.41 \\
G4$^{e}$     & $19^{h} 44^{m} 58.34^{s}$ &\ $-14^{h} 48^{m}  3.57^{s}$ & 78.9  & 0.12 & 9.9  & 0.06  & 0.48 \\
MS+RSG       & $19^{h} 44^{m} 56.73^{s}$ &\ $-14^{h} 48^{m} 50.20^{s}$ & -28.7 & 0.36 & 6.3  & 0.46  & 1.26 \\
RC           & $19^{h} 44^{m} 58.80^{s}$ &\ $-14^{h} 48^{m} 36.00^{s}$ & 67.1  & 0.22 & 11.8 & -0.06 & 1.02 \\
TRGB         & $19^{h} 44^{m} 57.63^{s}$ &\ $-14^{h} 48^{m} 12.62^{s}$ & 82.4  & 0.12 & 13.9 & 0.23  & 0.63 \\
RGB2         & $19^{h} 44^{m} 58.15^{s}$ &\ $-14^{h} 48^{m}  0.76^{s}$ & 71.3  & 0.12 & 10.1 & 0.10  & 0.43 \\
O-rich       & $19^{h} 44^{m} 56.13^{s}$ &\ $-14^{h} 48^{m} 26.01^{s}$ & 89.4  & 0.13 & 12.4 & 0.61  & 0.85 \\
C-rich       & $19^{h} 44^{m} 54.45^{s}$ &\ $-14^{h} 48^{m} 48.22^{s}$ & 104.3 & 0.08 & 4.9  & 1.03  & 1.22 \\
\enddata
\tablecomments{[$^{a}$] Estimate of galaxy center based on the 3D histogram of older stellar tracers 
(see Fig.~\ref{fig:center}).
[$^{b}$] Maximum likelihood performed over the entire sample of NGC~6822 candidate stars, but assuming the 
galaxy center based on the 3D histogram.
[$^{c}$] Maximum likelihood performed over the entire sample of NGC~6822 candidate stars, but keeping all 
the parameters free (see the top panel of Fig.~\ref{fig:GlobalRadial}).
[$^{d}$] Maximum likelihood performed by selecting candidate galaxy stars in magnitude 
(20.75 $< i <$ 23.5 mag) and in color ($g-i >$ 0.5 mag, see the middle panel of Fig.~\ref{fig:GlobalRadial}).
[$^{e}$] Maximum likelihood performed by selecting candidate galaxy stars in magnitude ($i<$23.5 mag, see 
the bottom panel of Fig.~\ref{fig:GlobalRadial}).}
\end{deluxetable*}
%_______________________________________________________________________________________

The bottom panels of the same figure display the radial distribution of old 
stellar tracers: TRGB (bottom left) and RGB2 (bottom right). We decided to 
investigate the radial distribution along the RGB, because plain physical arguments 
outlined in Section~\ref{sec:tracers} suggest that the role of old stellar populations 
increases when moving from fainter (RGB2) to brighter (TRGB) RGs. 
The solution of the fit for TRGB stars shows an almost spherical distribution
($\epsilon \sim$ 0.12) with a position angle of $\theta \sim$ 82 degrees. 
This means that old stellar tracers in this galaxy are almost edge-on with the 
semi-major axis along the E-W directions and the minor axis along the N-S 
direction. Moreover and even more importantly, the half-light radius is larger than 
for young and intermediate-age stellar populations and the isocontours cover the 
entire area covered by our optical photometry. This means that current photometry is 
not approaching the truncation radius of the galaxy. The radial distribution 
of the RGB2 plotted in the bottom right panel display properties that are 
intermediate between TRGB and RC stellar tracers. Indeed, their eccentricity 
is almost spherical and similar to old tracers, while the position angle is 
more similar to the intermediate stellar populations. 

AGB stars were closely investigated and characterized, this is the reason why we 
decide to focus our attention on their spatial distributions (Fig.~\ref{fig:OeCradial}). 
Note that the C-rich stars come from the selection performed on the 
\textit{r-K}--\textit{r-i} CCD because they are the largest sample when compared with the selection performed on the other selected CCDs (see Section~\ref{sec:ratio}). 
The O-rich star sample used here was obtained from our global optical selection 
of AGB stars (\textit{i}, \textit{g-i} CMD) once C-rich stars were removed. 
As expected, the M-type stars (left panel) show an almost spherical 
distribution and their structural parameters are very 
similar to the structural parameters of the old stellar tracers (bottom panels
of Fig.~\ref{fig:radial}). The similarity applies not only to the position angle and the 
eccentricity, but also to half-light radius. Thus suggesting that M-type stars 
are dominated by old- and intermediate-age progenitors. On the other hand, 
C-rich stars (right panel) are centrally concentrated and the peak of their distribution 
is off-center by 1.03 arcmin (RA) and 1.22 arcmin (DEC). This means that their 
distribution resemble the properties of the young stellar tracers 
(R$_{half}$ = 4.9$'$ vs 6.3$'$). However, they are spherically distributed 
($\epsilon \sim$ 0.08) as the old tracers, thus suggesting the key role of 
low mass stars for this stellar tracer.
The results of the radial distributions for O- and C-rich stars are consistent with 
those obtained by \citet{Hirsc20}.

%%%%%%%%%%%%%%%%%%%%%%%%%%%%%%%%%%%%%%%%%%%%%%%%%%%%%%%%%%%%%%%%%%%%%%%%%%%%%%%%%%%%%
\section{Population ratio between C- and O-rich stars} \label{sec:ratio}
%%%%%%%%%%%%%%%%%%%%%%%%%%%%%%%%%%%%%%%%%%%%%%%%%%%%%%%%%%%%%%%%%%%%%%%%%%%%%%%%%%%%%

The ratio between C- and M-type stars in a stellar population is used as a 
population diagnostic and to estimate the metallicity of the environment from which 
AGB stars formed. The reason why the C/M ratio is quite important for a quantitative 
analysis of resolved stellar populations is twofold. 
i) During the AGB phase, low and intermediate-mass stellar structures experience 
several physical mechanisms (3rd dredge-up, convective overshooting, mass loss, 
hot bottom burning) affecting the surface chemical composition. The efficiency 
of these mechanisms depends on several physical parameters, and in particular, on 
the stellar mass and on the chemical composition. Metal-poor and metal-intermediate 
stellar systems are fundamental laboratories to trace the transition from M- to 
C-type stars. The reader interested in a more quantitative discussion is referred to   
\citet{Weiss09}.  
ii) AGB stars are among the most important contributors to the integrated light 
of a galaxy \citep{Renzini86}. Their role becomes even more relevant 
in the NIR regime where the effects of dust obscuration is significantly reduced 
compared to the optical bands. Moreover and even more importantly, AGB stars 
play a key role in the chemical enrichment of galaxies, since they produce 
CNO and neutron capture (s-process) elements.  

\begin{figure}[ht!]
\centering
\includegraphics[width=8cm]{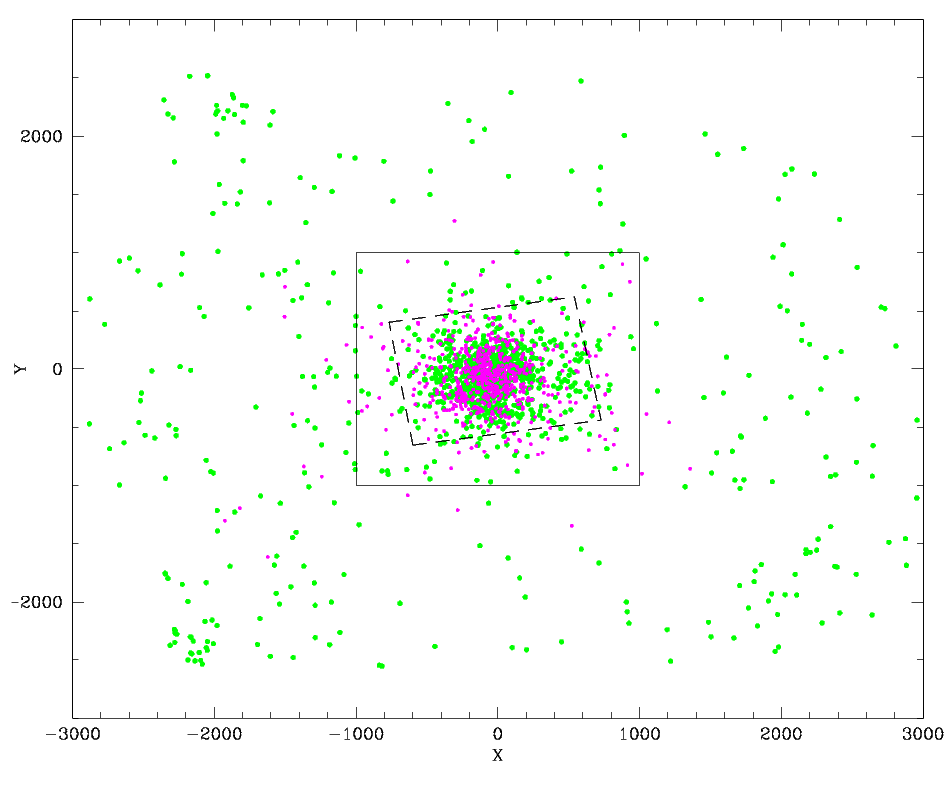}
\caption{Radial distribution of the C- (magenta dots) and O-rich (green dots) 
samples selected from the \textit{r-K}--\textit{r-i} CCD. 
The black dashed rectangle delimits the field 
of view of the Spitzer Space Telescope; the black box of 1000 square arcsec
outlines the area within which the star counts were made. Note that most of 
the C- and O-rich stars are located in a galaxy region entirely covered by the 
Spitzer dataset.
\label{fig:Cmap}}
\end{figure}

We paid special care in the selection of C- and O-rich stars and the C/M ratios 
based on the star counts performed on a wide range of CCDs are listed in 
Table~\ref{tab:AGBcount}.  Note that the table includes C/M ratios 
based on CCDs that were not discussed in Section~\ref{sec:AGB} because they 
display features similar to the CCDs showed in Figs.~\ref{fig:AGBcolor} and \ref{fig:Oseparation}. 
They were included to further constrain possible systematics in the use of CCDs 
based on different optical/NIR/MIR photometric bands.  
The radial distributions of candidate C- and M-type stars discussed in Section~\ref{sec:distribution}
show that the former group is significantly more centrally concentrated than 
the latter one. Moreover, optical, NIR and MIR datsets cover different 
galaxy regions (see Section~\ref{sec:data}). This means that star counts of AGB stars based 
on optical/NIR/MIR CCDs are sampling different regions of the galaxy. 
Fig.~\ref{fig:Cmap} shows the radial distributions of candidate C- and M-type 
stars selected on the \textit{r-K}--\textit{r-i} CCD, together with the sky 
area covered by MIR images collected with Spitzer (dashed black box). 
To provide homogeneous and accurate star counts of AGB stars we performed 
the selection over a box of 1000 square arcsec located across the center of the 
galaxy (solid black box). 

%________________________________________________________________________________
%________________________________________________________________________________

The number of C- and M-type selected by using optical/NIR/MIR CCDs and the 
C/M ratios are listed in Table~\ref{tab:AGBcount}. The errors on the C/M ratios take account 
for the Poisson uncertainties. The population ratios listed in this table show 
that the C/M ratio remain consistent, within the errors, when moving from selections 
based either on NIR (\textit{J-K}--\textit{H-K}, C/M=0.51 $\pm$ 0.04) or on 
optical/NIR (\textit{J-K}--\textit{g-i}, C/M=0.52 $\pm$ 0.04) to a selection 
based on narrow and broad-band photometry 
($CN-TiO$---$R-I$, C/M=0.49 $\pm$ 0.04)\footnote{Note that in 
the estimate of the C/M ratio we did not include a sample of 69 fainter/bluer 
carbon stars, as identified by \citetalias{Letarte02} (see Section~\ref{sec:AGB}). 
They deserve a more detailed spectroscopic analysis.}.   
However, the bulk of the selections based on optical/NIR CCDs provide an average 
C/M ratio of 0.68 $\pm$ 0.12, while those based on optical/MIR CCDs provide on average a 
C/M ratio of 0.63 $\pm$ 0.04, while optical/NIR/MIR CCDs give C/M ratios of 0.70 $\pm$ 0.08.    
The optical and the NIR datasets adopted in this investigation cover a significant 
fraction of the body of the galaxy. Therefore we decided to investigate the variation 
of the C/M ratio when using selections based on optical/NIR CCDs. As expected the 
population ratios estimated over the entire body of the galaxy are on average 20\% 
smaller than those limited to the innermost cluster regions. The difference is 
due to the fact that C-rich stars are centrally concentrated, while the O-rich 
stars cover a broader regions and their number steadily increases. This difference 
takes account for a significant fraction of the differences in the population ratios 
available in the literature (see Table~\ref{tab:literature}). 

%________________________________________________________________________________
\begin{deluxetable*}{lccccc}
%\tabletypesize{\scriptsize}
\tablenum{5}
\tablecaption{C/M population ratios based on a variety of optical/NIR/MIR color-color 
diagrams and iron abundances based on empirical C/M ratio-metallicity relations.  
\label{tab:AGBcount}}
\tablewidth{0pt}
\tablehead{
\colhead{CCD} & \colhead{C-rich} & \colhead{O-rich} & \colhead{C/M} & \colhead{[Fe/H]$^{a}$} & \colhead{[Fe/H]$^{b}$}
} %\decimalcolnumbers
\startdata
\multicolumn{6}{c}{} \\
\multicolumn{6}{c}{C-rich and O-rich star counts over a central area of 1000 square arcsec} \\
\textit{CN-TiO/R-I}      & 486    & 1002 & 0.49 $\pm$ 0.04 & -1.14 $\pm$ 0.10 & -1.24 $\pm$ 0.09 \\ 
\textit{J-K/g-i}         & 467    & 903  & 0.52 $\pm$ 0.04 & -1.15 $\pm$ 0.10 & -1.26 $\pm$ 0.09 \\ 
\textit{J-[3.6]/g-i}     & 454    & 743  & 0.61 $\pm$ 0.05 & -1.19 $\pm$ 0.09 & -1.29 $\pm$ 0.09 \\
\textit{J-K/H-K}         & 460    & 910  & 0.51 $\pm$ 0.04 & -1.15 $\pm$ 0.10 & -1.25 $\pm$ 0.09 \\ 
\textit{J-[3.6]/H-[3.6]} & 461    & 736  & 0.63 $\pm$ 0.05 & -1.20 $\pm$ 0.09 & -1.30 $\pm$ 0.08 \\		     
\textit{i-K/g-i}         & 546    & 824  & 0.66 $\pm$ 0.05 & -1.21 $\pm$ 0.09 & -1.31 $\pm$ 0.08 \\ 
\textit{r-K/r-i}         & 605    & 764  & 0.80 $\pm$ 0.06 & -1.26 $\pm$ 0.08 & -1.34 $\pm$ 0.07 \\ 
\textit{i-K/r-i}         & 588    & 780  & 0.75 $\pm$ 0.06 & -1.25 $\pm$ 0.08 & -1.33 $\pm$ 0.08 \\ 
\textit{i-[3.6]/g-i}     & 755    & 1053 & 0.70 $\pm$ 0.05 & -1.23 $\pm$ 0.09 & -1.32 $\pm$ 0.08 \\		     
\textit{i-[4.5]/g-i}     & 662    & 1145 & 0.58 $\pm$ 0.04 & -1.18 $\pm$ 0.09 & -1.28 $\pm$ 0.09 \\		    
\textit{r-[3.6]/r-i}     & 701    & 1107 & 0.63 $\pm$ 0.04 & -1.20 $\pm$ 0.09 & -1.30 $\pm$ 0.08 \\
\textit{i-[3.6]/r-i}     & 714    & 1095 & 0.65 $\pm$ 0.04 & -1.21 $\pm$ 0.09 & -1.30 $\pm$ 0.08 \\
\textit{r-[4.5]/r-i}     & 680    & 1127 & 0.60 $\pm$ 0.04 & -1.19 $\pm$ 0.09 & -1.29 $\pm$ 0.09 \\
\textit{i-[4.5]/r-i}     & 686    & 1121 & 0.61 $\pm$ 0.04 & -1.19 $\pm$ 0.09 & -1.29 $\pm$ 0.08 \\
\textit{i-[3.6]/r-J}     & 520    & 677  & 0.77 $\pm$ 0.06 & -1.25 $\pm$ 0.08 & -1.34 $\pm$ 0.07 \\
\textit{i-[4.5]/r-J}     & 494    & 703  & 0.70 $\pm$ 0.06 & -1.23 $\pm$ 0.09 & -1.32 $\pm$ 0.08 \\
\hline
\multicolumn{6}{c}{} \\
\multicolumn{6}{c}{C-rich and O-rich star counts made considering the entire area covered by our dataset} \\
\multicolumn{6}{c}{} \\
\textit{J-K/g-i}         & 482    & 1186 & 0.41 $\pm$ 0.03 & -1.09 $\pm$ 0.11 & -1.21 $\pm$ 0.10 \\
\textit{J-K/H-K}         & 474    & 1193 & 0.40 $\pm$ 0.03 & -1.09 $\pm$ 0.11 & -1.20 $\pm$ 0.10 \\
\textit{i-K/g-i}         & 561    & 1106 & 0.51 $\pm$ 0.04 & -1.15 $\pm$ 0.10 & -1.25 $\pm$ 0.09 \\
\textit{r-K/r-i}         & 624    & 1041 & 0.60 $\pm$ 0.04 & -1.19 $\pm$ 0.09 & -1.29 $\pm$ 0.09 \\
\textit{i-K/r-i}         & 607    & 1058 & 0.57 $\pm$ 0.04 & -1.18 $\pm$ 0.09 & -1.28 $\pm$ 0.09 \\
\enddata
\tablecomments{Iron abundances are based on the current C/M population ratios and the empirical 
relations provided by \citet{Battinelli05}[$^{a}$] and by \citet{Cioni09} [$^{b}$].}
\end{deluxetable*}
%_______________________________________________________________________________________

We also decided to use the new population ratios to provide a preliminary 
estimate of the mean metallicity of AGB stars in NGC~6822. Analytical relations 
correlating the population ratio to the mean iron abundance were provided by 
\citet{Battinelli05} by using a sample of LG galaxies: 
[Fe/H]$=-1.32(\pm 0.07) - 0.59(\pm 0.09) \times log(C/M)$.
The same relation was revised by \citet{Cioni09}, updating metallicity values 
by using RGB colour method. They found: 
[Fe/H]$=-1.39(\pm 0.06) - 0.47(\pm 0.10) \times log(C/M)$. 

We used these relations and the CM ratios listed in Table~\ref{tab:AGBcount} to 
estimate mean metallicity for the selected AGB stars, and the new estimated 
iron abundances are listed in columns 5 and 6 of the same table. The errors on the 
iron abundances take account for the uncertainties on both the coefficients 
of the analytical relations and the CM ratios. 
The iron abundances listed in Table~\ref{tab:AGBcount} indicate that AGB stars 
in NGC~6822 are independent of the adopted calibration, metal-intermediate, 
indeed we found a mean iron abundance of 
[Fe/H] $\sim$ -1.20 ($\sigma$ = 0.04 dex) and of [Fe/H] $\sim$ -1.30 ($\sigma$ = 0.03 dex), 
respectively. We note that the small values of the standard deviations is a 
consequence of the mild dependence of the adopted population ratio-metallicity relations on 
the C/M ratio.

Finally, we compared our results on population ratios and iron abundances with similar estimates available 
in the literature that are listed in Table~\ref{tab:literature}. C/M ratios are quite different from each other and from our average values, with the exception of the estimate from \citet{Hirsc20} (0.599 $\pm$ 0.036, based on NIR/MIR CMDs) that is very similar, within the errors, to our average C/M ratio (0.63 $\pm$ 0.04) based on optical/MIR CCDs. On the other side, the mean iron abundances are in good agreement with the literature estimates, except for the value from \citet{Kang06}. 
Metallicity measurements for NGC~6822 were also obtained by \citet{Davidge03} by using the slope of the RGB
([Fe/H] = -1 $\pm$ 0.3), and from spectroscopic measurements (\citealt{Venn01}: [Fe/H] = -0.49 $\pm$ 0.22;
\citealt{Tolstoy01}: [Fe/H] = -1 $\pm$ 0.5; \citealt{Kirby13}: [Fe/H] = -1.05 $\pm$ 0.01). 
Our estimates agree quite well with spectroscopic measurements of old 
stellar populations.

%_______________________________________________________________________________________
\begin{deluxetable}{cccccc}
\tablenum{6}
\tablecaption{C/M ratios and iron abundances available in literature for AGB stars in NGC~6822.\label{tab:literature}}
\tablewidth{0pt}
\tablehead{
\colhead{C/M} & & & \colhead{[Fe/H]} & & \colhead{Ref.$^{a}$}
} %\decimalcolnumbers
\startdata
1.0 $\pm$ 0.2	      & & & $\ldots$ 	           & & [1] \\
0.27 $\pm$ 0.03       & & & -0.99 $\pm$ 0.02       & & [2] \\
$\sim$ 1.05	          & & & -1.3 $\pm$ 0.2         & & [3] \\
0.48 $\pm$ 0.02$^{b}$ & & & -1.24 $\pm$ 0.07$^{b}$ & & [4] \\
0.29 $\pm$ 0.01$^{c}$ & & & -1.14 $\pm$ 0.08$^{c}$ & & [4] \\
0.95 $\pm$ 0.04       & & & -1.38 $\pm$ 0.06       & & [5] \\
0.599 $\pm$ 0.036     & & & -1.286 $\pm$ 0.095     & & [6] \\
\enddata
\tablecomments{$^{a}$ [1] \citetalias{Letarte02}; [2] \citet{Kang06}; [3] \citetalias{Kacharov12}; [4] \citetalias{Sibbons12}; [5] \citetalias{Sibbons15}; [6] \citet{Hirsc20}. \\
$^{b}$ C/M ratio and metallicity estimated over an area of 4 Kpc across the center of NGC~6822. \\
$^{c}$ C/M ratio and metallicity estimated over the entire observed area.}
\end{deluxetable}
%_______________________________________________________________________________________

%________________________________________________________________________________
\section{Summary and final remarks} \label{sec:summary}
%________________________________________________________________________________

We performed a detailed optical photometric analysis of the nearby dIrr galaxy 
NGC~6822. We carried out PSF photometry of a large set of $g$ (61), $r$ (18) and 
$i$-band (27) images collected with the wide field imager HSC available at the 
Subaru Telescope. The individual images have a FoV of 1.5 degrees in diameter 
and they include 104 CCDs. The dataset includes both shallow ($t\sim$ 30 s) 
and deep ($t\sim$ 240--300 s) exposures 
collected in very good seeing conditions ($\le$ 0.9$''$ for 50\% of the images).     
These images were complemented with a 13 $g,r,i$ images collected with the 
wide field imager MegaPrime available at CFHT, with two dozen $g,r,i$  
images collected with DECAM available at CTIO 4m Blanco telescope and with
6~$g$-band images acquired with the WFC available at INT. These 
data cover the same sky area covered by the SUBARU dataset, and they 
have exposure times ranging from 15 to 600 $s$ and seeing similar to the HSC images. 
The individual images cover a FoV of 1x1 square degree (MegaPrime), of 2.2 degrees 
in diameter (DECAM) and of 34$\times$34 arcmin (WFC), and they were included to 
improve the absolute photometric calibration and the sampling of bright stars.  
All in all the current photometric dataset covers an area of 2 square degrees 
across the galaxy center with three different photometric bands. We performed 
PSF photometry over 7115 individual CCD images by using the suits of 
codes DAOPHOT-ALLSTAR-ALLFRAME. The final catalog includes more than 1 million 
stars with at least one measurement in two different photometric bands. The 
maximum number of measurements per star range from 74 ($g$), to 28 ($r$) and 
to 42 ($i$). In total, in the three photometric bands we performed $4 \times 10^7$ 
photometric measurements of objects across the entire field of view. The current 
catalog is the widest and most homogeneous photometric dataset ever collected 
for a nearby dIrr, but the Magellanic Clouds. The limiting magnitudes range from 
14 to 25.8 mag ($\sigma$=0.04 mag) in the $g$-band, the deepest, 
from 13.7 to 25.5 mag ($\sigma$=0.05 mag) in the $r$-band, and from 
13 to 24.9 mag ($\sigma$=0.09 mag) in the $i$-band, the shallowest.    

To provide a detailed analysis of stellar populations across the entire FoV 
of the current dataset, we developed a new approach to identify candidate 
field and galaxy stars. We took advantage of the three different photometric 
bands and the new algorithm is an iterative procedure based on the 3D 
CCM diagram. We identified a training set of candidate 
galaxy stars located at a few arcmin from the galaxy center and the key 
evolutionary sequences were subtracted from the global 3D CCM diagram.  
After two iterations, we ended up with a catalog of candidate galaxy stars 
including more than 550,000 stars with at least one measurement in two different 
photometric bands. 

Finally, the global optical photometric catalog was cross correlated with the 
NIR (\citetalias{Sibbons12}) and MIR \citep{Khan15, Marocco21} catalogs available in the 
literature. The NIR catalog includes $JHK$ photometry and it is based on the
UKIRT images covering almost the same sky area of our optical dataset, 
while the MIR catalogs include both the IRAC photometry from images 
collected with SPITZER and covering the innermost regions of the galaxy, and 
the CatWISE and AllWISE photometry obtained from the WISE survey and
covering an area of 3$\times$3~degrees across the center of galaxy.   
The global optical-NIR-MIR catalog includes more than 65,000 stars with 
at least two measurements in optical, in NIR and in MIR photometric bands.  

The global optical-NIR-MIR catalog has been the stepping stone for several 
interesting results concerning the stellar content of NGC~6822. \\

(i) We estimated apparent and absolute total magnitudes in the nine 
photometric bands adopted in the current investigations. The absolute 
magnitude range from $\sim$ -15 to -16 mag in the three optical ($g,r,i$) bands, 
from $\sim$ -15.5 to -16.5 mag in the three NIR ($J,H,K$) bands,
and from $\sim$ -17.6 to -19.3 mag in the three MIR ($W1,W2,W3$) bands. 
We also compared the current estimates with the absolute magnitudes of the MCs, 
that were estimated from the Gaia eDR3 catalogue, suggesting that NGC~6822 is 
fainter in the optical regime than the MCs. However, optical-MIR 
colors suggest that NGC~6822 is systematically redder than MCs. The difference 
is mainly caused by the high number of C-rich stars identified in the former system.  
Furthermore, we found that NGC~6822 in the MIR color-color diagram 
($W1-W2$, $W2-W3$; \citealt{Hain16}) is located in the blue tail of the 
so-called ``composite" stellar systems \citep{Baldwin81, Kewley01, Kauffmann03}, 
i.e. dwarf galaxies in which the emission-line flux includes contributions 
from both AGN and star formation activity. Thus suggesting that this diagnostic
is also prone to possible systematics introduced by AGB stars. 

(ii) We performed a new and independent estimate of the center of galaxy 
by using two different approaches: 3D histogram and maximum likelihood. 
The 3D histogram method was applied to the old stellar tracers, and we 
found that the coordinates of the galaxy center are: 
R.A.(J2000) = $19^{h} 44^{m} 58^{s}.56$, DEC(J2000) = $-14^{d} 47^{m} 34^{s}.8$. 
We also used the maximum likelihood algorithm from \citet{Martin08} to estimate 
structural parameters and the coordinates of the galaxy center agree, within the 
errors, with the 3D histogram. The difference with previous estimates available 
in the literature is of the order of 1.15 arcmin in RA and 1.53 arcim in DEC.  
The spatial distribution and the isocontours of the whole catalog show that the 
galaxy is almost spherically and symmetrically distributed and includes stars 
at radial distances larger than 1 degree. This evidence indicates that the 
truncation radius of this galaxy is well beyond the current estimates. Moreover, 
the solution of the fit provided a position angle of $\theta \sim$ 75 degrees 
and a very modest eccentricity ($\epsilon \sim$ 0.15) that are in quite fair agreement 
with similar estimates available in literature (\citealt{Battinelli06}, \citealt{Zhang21}).

(iii) The reduction strategy we devised provided deep and accurate CMDs. The 
\textit{i}, \textit{g-i} CMD of candidate galaxy stars (Fig.~\ref{fig:optical}) 
covers more than nine $i$-band magnitudes. This CMD shows several
well-defined evolutionary sequences (Young MS, RSG, RGB, RC, AGB) suggesting 
the presence of young, intermediate-age and old stellar populations in the galaxy.
We selected the different stellar tracers on the optical CMD and in order to 
overcome possible systematics due to the differential reddening affecting this galaxy, 
we validated our selections by using the Wesenheit index and by taking advantage 
of both NIR and MIR photometry. 
Special attention was paid to the identification of AGB stars. We adopted the 
the same approaches suggested in literature (\citetalias{Letarte02}, \citealt{Kang06}) 
and investigated new optical-NIR-MIR color combinations to further improve the 
identification and characterization of C- and O-rich stars.
We found that the \textit{CN-TiO}, \textit{R-I} CCD (defined by \citetalias{Letarte02})
and a series of new optical-NIR, optical-MIR and optical-NIR-MIR CCDs are very solid diagnostics 
to identify both C-rich and O-rich stars. The former one was already known in the literature, 
but the region covered by M-type stars is affected by field star contamination. The latter ones 
are very promising because C- and O-rich stars are distributed along either two almost parallel 
sequences or along two sequences with different slopes.  
In passing we also note that the optical-NIR \textit{r-K}--\textit{r-i} and \textit{i-K}--\textit{r-i} 
CCDs display a separation in two parallel sequences of candidate O-rich stars, suggesting that
they might include a younger (bluer sequence) and an older (redder sequence) AGB component.
We compared the slopes of the sequences defined by candidate C- and O-rich
in the different CCDs with the slopes of the respective reddening vectors, finding that
they all differ from each other. This means that the dispersion in color present in 
the adopted CCDs is an intrinsic feature of AGB stars and it is minimally affected 
by differential reddening. Moreover, the current identifications of C- and O-rich samples 
are fully supported by the position of the spectroscopically confirmed C- and M-type stars 
provided by \citetalias{Kacharov12} and by \citetalias{Sibbons15}.

(iv) We investigated the radial distributions and provided the structural parameters 
for the different stellar tracers selected in the optical CMD. 
The results show that young stellar tracers (MS+RSG) are distributed along 
a well-defined bar that is off-center by 0.46 arcmin in RA and 1.26 arcmin in DEC; 
they have a modest half-light radius (R$_{half} \sim$ 6.3$'$) and the young stars located outside the bar
suggest the possible presence of a disk ($\theta \sim$ -29 degrees). Indeed, their isocontours
seems to follow very well the isocontours traced by the HI column density map 
\citep{deBlok00, deBlok03, deBlok06}.
The intermediate-age stellar tracers (RC) display isodensity contours that are far from being symmetric,
they are distributed over a significant fraction of the body of the galaxy (R$_{half}$ = 11.8$'$) and
have a smaller eccentricity when compared with young stellar tracers ($\epsilon \sim$ 0.2 vs 0.4).
The structural properties of the two selected old stellar tracers, TRGB and RGB2 stars, differ 
from each other. The TRGB stars show an almost spherical distribution
($\epsilon \sim$ 0.1), are almost edge-on with the semi-major axis along the 
E-W directions and the minor axis along the N-S direction ($\theta \sim$ 82 degrees)
and have the largest half-light radius among the other stellar populations (R$_{half} \sim$ 14$'$). 
On the other hand, the RGB2 stars display properties that are similar to both TRGB and 
RC stellar tracers ($\epsilon \sim$ 0.1, $\theta \sim$ 71 degrees and R$_{half} \sim$ 10$'$).
Moreover, we also found that O- and C-rich stars (AGB) have different properties.
The O-rich stars have structural parameters very similar to that of the old tracers
($\epsilon \sim$ 0.1, $\theta \sim$ 89 degrees and R$_{half}$ = 12.4$'$), thus indicating
that they are dominated by old- and intermediate-age progenitors. On the contrary,
the C-rich stars are more centrally concentrated and have structural parameters similar 
to both young, indeed the peak of the distribution is off-center and R$_{half}$ = 4.9$'$,
and old stellar tracer ($\epsilon \sim$ 0.1). 

(v) We estimated the C/M population ratios based on the star counts performed on the 
selected CCDs, obtaining on average a C/M ratio of 0.68$\pm$0.12 for optical/NIR CCDs,
a C/M of 0.63$\pm$0.04 for optical/MIR CCDs, a C/M of 0.70$\pm$0.08 for optical/NIR/MIR CCDs
and a C/M of 0.49 $\pm$ 0.04 for the narrow and broad-band photometry (\textit{CN-TiO}--\textit{R-I}) CCD. 
We also provided an estimate of the mean metallicity of AGB stars by using the current 
C/M population ratios and the empirical relations provided by \citet{Battinelli05} and 
by \citet{Cioni09}. We found mean iron abundances of [Fe/H] $\sim$ -1.20 ($\sigma$ = 0.04 dex)
and of [Fe/H] $\sim$ -1.30 ($\sigma$ = 0.03 dex), respectively. Mean metallicity values 
are in fair agreement with the literature estimates (e.g., \citetalias{Kacharov12}, 
\citetalias{Sibbons12}, \citealt{Hirsc20}, \citealt{Davidge03}, \citealt{Tolstoy01}) 
suggesting a metal-intermediate iron abundance.

The approach developed in the current investigation and the definition of new photometric 
diagnostics to investigate evolutionary properties of stellar populations in nearby stellar 
systems are a good viaticum for future optical-NIR-MIR photometric surveys. In this 
context the incoming optical survey by the Vera Rubin Observatory is going to play a key 
role, since the main survey plans to cover the entire southern sky every three nights 
in six different photometric bands ($u,g,r,i,z,y$) with limiting magnitudes ($\sim$27-28 mag) 
that will allow us a complete census of stellar populations older than 10 Gyrs (Main Sequence 
turn-off) over the entire Local Group.

%_______________________________________________________________________________________
%_______________________________________________________________________________________
\section*{Acknowledgements}

We are grateful to an anonymous referee for her/his positive words
concerning the content and the cut of an early version of the current paper, and for
her/his very pertinent suggestions that improved its readability.

It is a real pleasure to thank P. Battinelli and S. Demers for sending
us their catalog of NGC~6822 in elettronic form. We are also very
grateful to W.J.G. de Blok for supplying his HI map for NGC 6822 in
elettronic form.

M. Monelli acknowledges financial support from the Spanish Ministry of Science
and Innovation (MICINN) through the Spanish State Research Agency, under
the grant PID2020-118778GB-I00
and under the Severo Ochoa Programe 2020-2023 (CEX2019-000920-S).
M. Marengo and J.P. Mullen are supported by the National Science Foundation 
under grant No. AST-1714534.
M. Salaris acknowledges support from The Science and Technology Facilities 
Council Consolidated Grant ST/V00087X/1.

This research has made use of the GaiaPortal catalogues access tool, 
ASI - Space Science Data Center, Rome, Italy (\url{http://gaiaportal.ssdc.asi.it}).

This paper is based on data collected at the Subaru Telescope and retrieved from the HSC data archive system, which is operated by the Subaru Telescope and Astronomy Data Center at NAOJ. Data analysis was in part carried out with the cooperation of Center for Computational Astrophysics (CfCA), NAOJ. The HSC collaboration includes the astronomical communities of Japan and Taiwan, and Princeton University. The HSC instrumentation and software were developed by the National Astronomical Observatory of Japan (NAOJ), the Kavli Institute for the Physics and Mathematics of the Universe (Kavli IPMU), the University of Tokyo, the High Energy Accelerator Research Organization (KEK), the Academia Sinica Institute for Astronomy and Astrophysics in Taiwan (ASIAA), and Princeton University. Funding was contributed by the FIRST program from the Japanese Cabinet Office, the Ministry of Education, Culture, Sports, Science and Technology (MEXT), the Japan Society for the Promotion of Science (JSPS), Japan Science and Technology Agency (JST), the Toray Science Foundation, NAOJ, Kavli IPMU, KEK, ASIAA, and Princeton University. 

The Pan-STARRS1 Surveys (PS1) and the PS1 public science archive have been made possible through contributions by the Institute for Astronomy, the University of Hawaii, the Pan-STARRS Project Office, the Max Planck Society and its participating institutes, the Max Planck Institute for Astronomy, Heidelberg, and the Max Planck Institute for Extraterrestrial Physics, Garching, The Johns Hopkins University, Durham University, the University of Edinburgh, the Queen’s University Belfast, the Harvard-Smithsonian Center for Astrophysics, the Las Cumbres Observatory Global Telescope Network Incorporated, the National Central University of Taiwan, the Space Telescope Science Institute, the National Aeronautics and Space Administration under grant No. NNX08AR22G issued through the Planetary Science Division of the NASA Science Mission Directorate, the National Science Foundation grant No. AST-1238877, the University of Maryland, Eotvos Lorand University (ELTE), the Los Alamos National Laboratory, and the Gordon and Betty Moore Foundation.

This work is based in part on observations made with the Spitzer Space
Telescope, which is operated by the Jet Propulsion Laboratory,
California Institute of Technology under a contract with the
National Aeronautics and Space Administration (NASA).

Some of the data reported here were obtained as part of the UKIRT Service Programme. UKIRT is owned by the University of Hawaii (UH) and operated by the UH Institute for Astronomy. When some of the data reported here were obtained, UKIRT was operated by the Joint Astronomy Centre on behalf of the Science and Technology Facilities Council of the U.K.

This publication makes use of data products from the Wide-field Infrared Survey Explorer, which is a joint project of the University of California, Los Angeles, and the Jet Propulsion Laboratory/California Institute of Technology, funded by the National Aeronautics and Space Administration. 

This publication makes use in part of data products from the Two Micron All Sky Survey, which is a joint project of the University of Massachusetts and the Infrared Processing and Analysis Center/California Institute of Technology, funded by the National Aeronautics and Space Administration and the National Science Foundation.

This work presents results from the European Space Agency (ESA) space mission Gaia. Gaia data are being processed by the Gaia Data Processing and Analysis Consortium (DPAC). Funding for the DPAC is provided by national institutions, in particular the institutions participating in the Gaia MultiLateral Agreement (MLA).

We made use of the VizieR catalog access tool, provided by CDS, Strasbourg, France (DOI: 10.26093/cds/vizier).

\textit{Software:} Astropy \citep{Astropy13}, Numpy \citep{Harris20}, SExtractor \citep{Bertin96}, Topcat \citep{Taylor05}

%_______________________________________________________________________________________
%_______________________________________________________________________________________
\appendix

\section{Identification of the peak associated to RC stars} \label{appendix1}

\begin{figure}[htbp!]
\centering
\includegraphics[width=11cm]{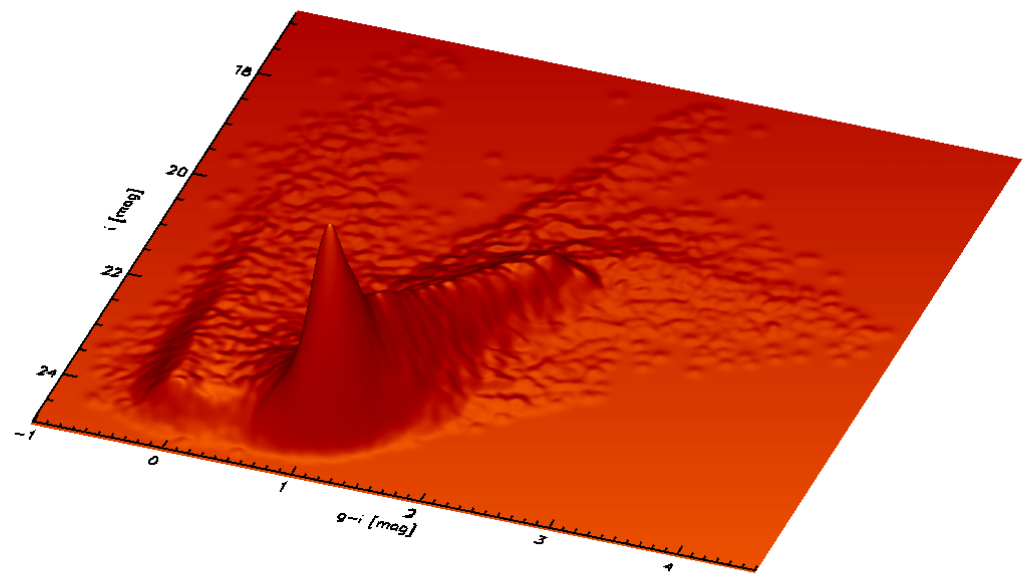}
\caption{3D histogram of the $i$, $g-i$ CMD. The view is from the faint to the 
brighter limit of the CMD. Note that in this CMD we are plotting all the stars 
with at least one measurement in $g$ and $i$ bands. The Z-axis is in arbitrary 
units. From bluer to redder colors the relatively shallow sequences are associated 
to young MS, to RSG and to AGB stars. The steady increase in the star counts present 
in the central regions is associated to RG stars. The main peak is associated to RC stars.  
\label{fig:3dcmd}}
\end{figure}

We already mentioned in section~\ref{sec:tracers} that RC stars are solid tracers of intermediate-age 
stellar tracers. However, the identification of the peak associated to RC stars is far 
from being trivial, in a stellar system like NGC~6822, because the magnitudes and colors 
of RC stars in optical CMDs overlap with RGB stars. To overcome this limitation we 
adopted the same approach suggested by \citet{Sanna08} for investigating stellar 
populations in IC10. Fig.~\ref{fig:3dcmd} shows the 3D histogram of a $i$, $g-i$ CMD. The CMD was 
split with a grid in magnitude and color and the Z-axis shows the number of stars per 
grid point. The number of stars was smoothed with a Gaussian kernel with an unitary weight 
and $\sigma$ equal to the Poisson error associated to the star count. We performed a 
series of test and trials to properly identify the optimal bin size in magnitude 
and in color and Fig.~\ref{fig:3dcmd} shows the final outcome. Data plotted in this figure display, 
from left to right, several well defined, but relatively shallow, evolutionary sequences, 
namely the young MS, the RSG and the AGB stars. 
The steady increase in star counts located 
in the central region of the CMD is associated, from top to bottom, to TRGB and RGB stars.
A preliminary comparison with evolutionary models suggests that the main peak located 
at $i \sim$ 24 mag and $g-i \sim$ 1 mag is caused by RC stars. A more detailed discussion concerning the 
comparison between theory and observations will be addressed in a forthcoming paper 
(Tantalo et al. 2022, in preparation).     

%________________________________________________________________________________
%________________________________________________________________________________

\begin{figure}[htbp!]
\centering
\includegraphics[width=8cm]{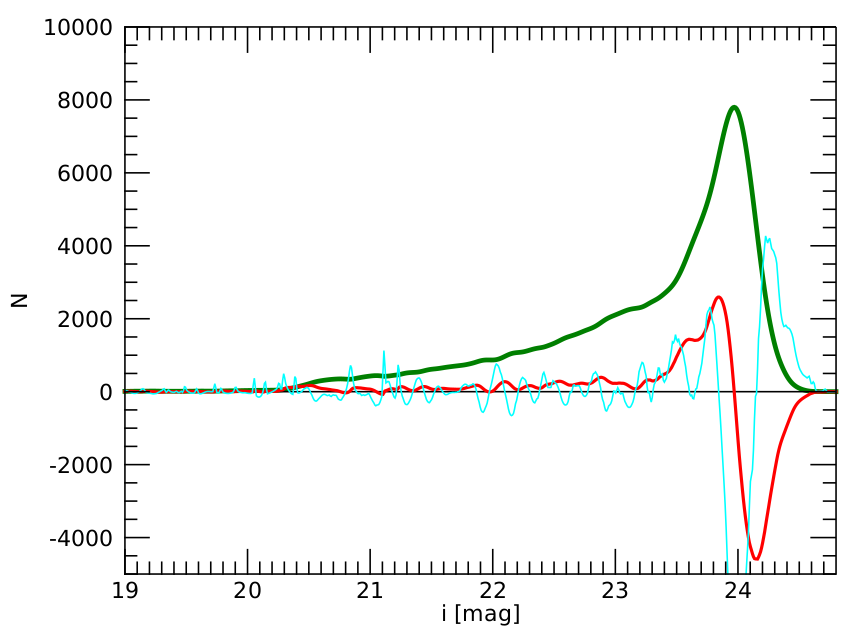}
\caption{Ridge line in magnitude ($i$-band, green line) of the star counts 
associated to RG stars showed in Fig.~\ref{fig:3dcmd}. The red and cyan lines show, 
respectively, the first and the second derivative of the ridge line.   
\label{fig:ridge}}
\end{figure}

To constrain on a more quantitative basis the location of RC stars we isolated the star
counts associated to RG stars and estimated the ridge line in magnitude. Fig.~\ref{fig:ridge} shows 
the ridge line in magnitude (green line) and its first and second derivatives 
(red and cyan lines). The maximum in the ridge line and the minimum in the first 
derivative located at $i \sim$ 23.9--24.0 mag trace the RC stars.     

%________________________________________________________________________________
%________________________________________________________________________________

\section{Selection of candidate galaxy stars for the Magellanic Clouds} \label{appendix2}

The selections for the Magellanic Clouds were obtained by querying the 
\textit{Gaia} eDR3 database. In particular, we selected a sky area with a 20$^\circ$ 
radius around the centre of the LMC and with a sky area with 11$^\circ$ radius for 
the SMC by using the following selection criteria to control the quality of the data: 
not {\tt duplicated\_source}; with color {\tt bp\_rp} measured; with astrometric 
solutions either with 5- or 6-parameters; not QSO; {\tt ruwe}$<1.4$; $G<20.5$; 
$C^* < 1.0$. The $C^*$ is the corrected {\tt phot\_bp\_rp\_excess\_factor} 
introduced by \citet{Riello21} wich helps to remove background galaxies; 
bright QSOs can be identified by using the {\tt agn\_cross\_id} table included 
in the \textit{Gaia} release.

\begin{figure}[htbp!]
\centering
\includegraphics[width=16cm]{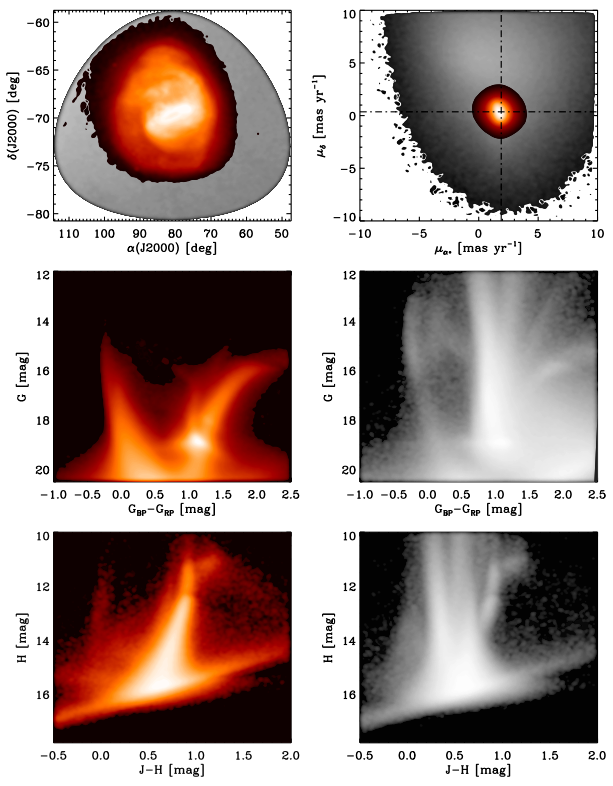}
\caption{Top: sky coverage (left) and proper motions (right) of the candidate 
LMC stars (coloured). The grey region highlights candidate field stars. 
Middle: Optical, $G,G_{BP}-G_{RP}$, CMD for candidate LMC (left) and field (right) stars. 
Bottom: Same as the middle, but for the NIR, $H,J-H$, CMD.
\label{fig:lmc}}
\end{figure}

\begin{figure}[htbp!]
\centering
\includegraphics[width=16cm]{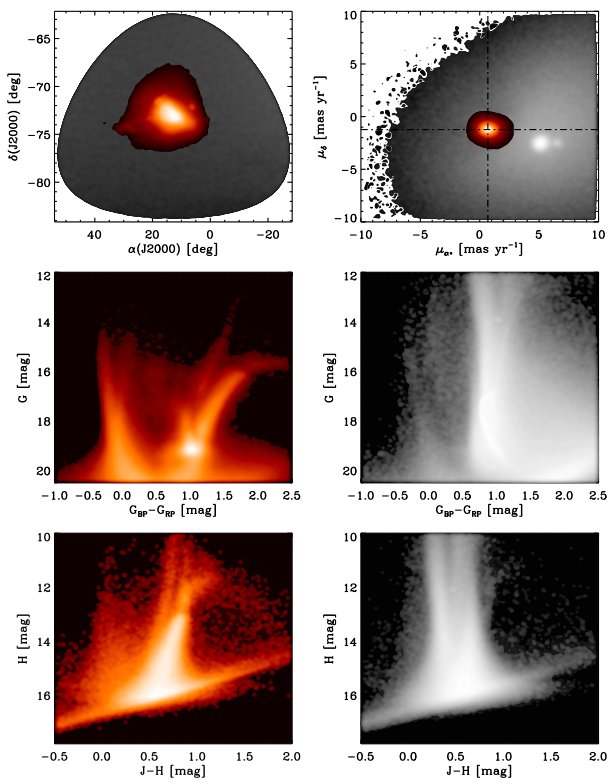}
\caption{Same as the Fig.~\ref{fig:lmc}, but for candidate SMC and field stars.   
\label{fig:smc}}
\end{figure}

We ended up with a sample including 4,134,870 objects for the SMC and 
14,716,026 for the LMC (see the top-left panels of Fig.~\ref{fig:lmc} and 
\ref{fig:smc}). The exquisite quality of \textit{Gaia} astrometric data 
allowed us to separate the stellar populations of the MCs from the foreground 
field stars by using the proper motion selection criteria. To determine the 
mean proper motion of both LMC and SMC we used sub-samples of objects that 
are closer than 0.5$^\circ$ to the centres of the galaxies and then we computed 
a Gaussian fit of the individual $\mu_{\alpha^*}$ and $\mu_\delta$ distributions. 
Therefore, we obtained the centres of proper motion at [0.68, -1.23]~mas~yr$^{-1}$ 
for the SMC and [1.89,0.36]~mas~yr$^{-1}$ for the LMC. Subsequently, we selected 
all the objects compatibles, within 10 times the errors, with the quoted mean 
proper motions (see the top-right panel of Fig.~\ref{fig:lmc} and \ref{fig:smc}).
Finally, we minimised the foreground contamination by selecting stars with 
{\tt parallax}$/${\tt parallax$_{error}$}$<$5. This parallax cut excludes solutions 
that are not compatible with being distant enough to be part of the LMC or SMC, 
and therefore possible foreground contamination from Milky Way stars. 
The final samples include 2,380,151 stars for the SMC and 12,456,270 for the LMC 
(see the middle panels of Fig.~\ref{fig:lmc} and \ref{fig:smc}).
We also took advantage of the pre-computed cross-matches of the 
\textit{Gaia} catalog with visual and infrared large surveys \citep{Marrese19}, 
in particular, with 2MASS (see the bottom panels of Fig.~\ref{fig:lmc} and \ref{fig:smc}) 
to further constrain the properties of candidate galaxy stars.

%________________________________________________________________________________
%________________________________________________________________________________

%________________________________________________________________________________
%________________________________________________________________________________

%_______________________________________________________________________________________
%_______________________________________________________________________________________

\bibliography{Paper_NGC6822}{}
\bibliographystyle{aasjournal}



\end{document}